\documentclass[aps,
pra,
twocolumn,
superscriptaddress,
nolongbibliography,
nofootinbib,
10pt,
]{revtex4-2}
\usepackage[utf8]{inputenc}
\usepackage{amsmath}
\usepackage{mathtools}
\usepackage{amssymb}
\usepackage{graphicx}
\usepackage{xcolor,soul}
\usepackage{siunitx}
\usepackage{dsfont}
\usepackage{bm}
\usepackage{braket}
\usepackage[
	hidelinks,
	colorlinks=true,
	linkcolor=black,
	citecolor=blue,
	urlcolor=black,
]{hyperref}\usepackage[english]{babel}
\usepackage{float}
\usepackage{etoolbox}

\usepackage[english,nomargin,inline,marginclue,draft]{fixme}
\fxusetheme{colorsig}
\FXRegisterAuthor{as}{asc}{\color{magenta}AS}
\FXRegisterAuthor{tc}{atc}{\color{purple}TC}
\FXRegisterAuthor{aw}{aaw}{\color{orange}AW}
\FXRegisterAuthor{lp}{alp}{\color{cyan}LP}
\FXRegisterAuthor{ib}{ifb}{\color{blue}IB}

\usepackage[todonotes={textsize=scriptsize},]{changes}
\definechangesauthor[name=Thomas, color=orange]{TC}
\definechangesauthor[name=Immanuel, color=blue]{IB}
\definechangesauthor[name=Antoine, color=purple]{AG}

\def\nobreakbefore{\relax\ifvmode\else
    \ifhmode
      \ifdim\lastskip > 0pt\relax
        \unskip\nobreakspace
      \else \nobreakspace
      \fi
    \fi
  \fi
}
\let\oldcite\cite
\renewcommand\cite{\nobreakbefore\oldcite}

\newcommand{\kB}{\ensuremath{k_{\rm B}}} 

\newcommand{\ee}{\ensuremath{\mathrm{e}}}
\newcommand{\ii}{\ensuremath{\mathrm{i}}}
\newcommand{\qafm}{\ensuremath{\bm{q}_{\mathrm{AFM}}}}
\newcommand{\dmax}{\ensuremath{d_{\mathrm{max}}}}
\newcommand{\Spp}{\ensuremath{\mathcal{S}_{\mathrm{AFM}}}}
\newcommand{\scaling}{\Theta}
\newcommand{\figref}[1]{Fig.~\ref{#1}}
\newcommand{\Eqref}[1]{Eq.~\eqref{#1}}
\newcommand{\secref}[1]{Sec.~\ref{#1}}
\DeclareSIUnit\Gauss{G}
\usepackage{booktabs}

\makeatletter
\def\maketitle{
\@author@finish
\title@column\titleblock@produce
\suppressfloats[t]}
\makeatother

\begin{document}

\newcommand{\MPQ}{Max-Planck-Institut f\"{u}r Quantenoptik, 85748 Garching, Germany}
\newcommand{\MCQST}{Munich Center for Quantum Science and Technology, 80799 Munich, Germany}
\newcommand{\LMU}{Fakult\"{a}t f\"{u}r Physik, Ludwig-Maximilians-Universit\"{a}t, 80799 Munich, Germany}
\newcommand{\LCF}{Laboratoire Charles Fabry, Institut d'Optique Graduate School, CNRS, Universit\'e Paris-Saclay, 91127 Palaiseau, France}
\newcommand{\CDF}{Coll{\`e}ge de France, PSL University, 11 place Marcelin Berthelot, 75005 Paris, France}
\newcommand{\CCQ}{Center for Computational Quantum Physics, Flatiron Institute, 162 5th Avenue, New York, New York 10010, USA}
\newcommand{\CPHT}{CPHT, CNRS, {\'E}cole Polytechnique, IP Paris, F-91128 Palaiseau, France}
\newcommand{\UNIGE}{DQMP, Universit{\'e} de Gen{\`e}ve, 24 quai Ernest Ansermet, CH-1211 Gen{\`e}ve, Suisse}
\newcommand{\MPIPKS}{Max Planck Institute for the Physics of Complex Systems, N\"othnitzer Strasse 38, Dresden 01187, Germany}
\newcommand{\ASC}{Arnold Sommerfeld Center for Theoretical Physics (ASC), Ludwig-Maximilians-Universit\"{a}t, 80799 Munich, Germany}
\newcommand{\UR}{University of Regensburg, Universit\"{a}tsstr. 31, 93053 Regensburg, Germany}
\newcommand{\LZU}{Lanzhou Center for Theoretical Physics, Key Laboratory of Quantum Theory and Applications of MoE, Key Laboratory of Theoretical Physics of Gansu Province, and School of Physical Science and Technology, Lanzhou University, Lanzhou, Gansu 730000, China.}
\newcommand{\STRATH}{Department of Physics and SUPA, University of Strathclyde, Glasgow G4 0NG, United Kingdom}

\title{
Observation of emergent scaling of spin-charge correlations\\at the onset of the pseudogap
}

\author{Thomas Chalopin}\email[Electronic address: ]{thomas.chalopin@institutoptique.fr}\affiliation{\MPQ}\affiliation{\MCQST}\affiliation{\LCF}

\author{Petar Bojovi\'c}
\affiliation{\MPQ}\affiliation{\MCQST}

\author{Si Wang}
\affiliation{\MPQ}\affiliation{\MCQST}

\author{Titus Franz}
\affiliation{\MPQ}\affiliation{\MCQST}

\author{Aritra Sinha}
\affiliation{\MPIPKS}

\author{Zhenjiu Wang}
\affiliation{\LMU}
\affiliation{\ASC}
\affiliation{\LZU}

\author{Dominik Bourgund}
\affiliation{\MPQ}\affiliation{\MCQST}

\author{Johannes Obermeyer}
\affiliation{\MPQ}\affiliation{\MCQST}

\author{Fabian Grusdt}
\affiliation{\MCQST}\affiliation{\LMU}\affiliation{\ASC}

\author{Annabelle Bohrdt}
\affiliation{\UR}
\affiliation{\MCQST}

\author{Lode Pollet}
\affiliation{\MCQST}\affiliation{\LMU}\affiliation{\ASC}

\author{Alexander Wietek}
\affiliation{\MPIPKS}

\author{Antoine Georges}
\affiliation{\CDF}
\affiliation{\CCQ}
\affiliation{\CPHT}
\affiliation{\UNIGE}

\author{Timon Hilker}
\affiliation{\MPQ}\affiliation{\MCQST}\affiliation{\STRATH}

\author{Immanuel Bloch}
\affiliation{\MPQ}\affiliation{\MCQST}\affiliation{\LMU}

\begin{abstract}
In strongly correlated materials, interacting electrons are entangled and form collective quantum states, resulting in rich low-temperature phase diagrams \cite{morosan:2012}.
Notable examples include cuprate superconductors, in which superconductivity emerges at low doping out of an unusual ``pseudogap'' metallic state above the critical temperature \cite{damascelli:2003, lee:2006, timusk:1999,norman:2005}.
The Fermi-Hubbard model \cite{hubbard:1963}, describing a wide range of phenomena associated with strong electron correlations \cite{dagotto:1994, qin:2022}, still offers major computational challenges despite its simple formulation.
In this context, ultracold atoms quantum simulators have provided invaluable insights into the microscopic nature of correlated quantum states \cite{mazurenko:2017a,koepsell:2019,hirthe:2023,bourgund:2025}.
Here, we use a quantum gas microscope Fermi-Hubbard simulator to explore a wide range of dopings and temperatures in a regime where a pseudogap is known to develop.
By measuring multi-point correlation functions up to fifth order, we uncover a novel universal scaling behaviour in magnetic and higher-order spin-charge correlations characterised by a doping-dependent temperature scale.
Accurate comparisons with determinant Quantum Monte Carlo and Minimally Entangled Typical Thermal States simulations confirm that this temperature scale is comparable to the pseudogap temperature $T^{*}$.
Our quantitative findings reveal a novel qualitative behaviour of magnetic properties and spin-charge correlations in an emergent pseudogap and pave the way towards the exploration of charge pairing and collective phenomena expected at lower temperatures.
\end{abstract}

\maketitle

\begin{figure}[!t]
\centering
\includegraphics[scale=1]{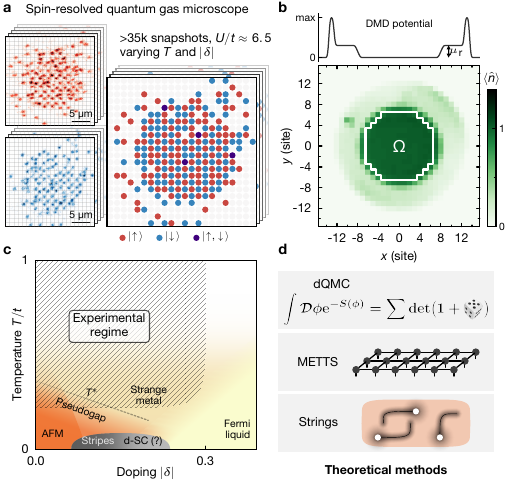}
\caption{
\textbf{Quantum simulation of the Fermi-Hubbard model.}
\textbf{a.} Examples of experimental snapshots using a quantum gas microscope with spin and charge resolution.
\textbf{b.} Averaged atomic density, depicting the central region ($\Omega$) over which the analysis is carried out, and the surrounding reservoir with chemical potential $\mu_\mathrm{r}$ experimentally adjusted using a DMD (see SI).
\textbf{c.} Conjectured phase diagram of the FHM.
The hatched region approximately depicts the regime accessed by our experimental apparatus.
AFM: region with sizeable antiferromagnetic correlations.
d-SC: conjectured superconducting phase.
\textbf{d.} Theoretical methods employed in this work: dQMC, METTS, and geometric strings (see text).
}
\label{fig:fig1}
\end{figure}

The Fermi-Hubbard model (FHM) is a minimal but at the same time paradigmatic model for describing interacting fermions on a lattice \cite{hubbard:1963}.
The conjectured phase diagram of the two-dimensional (2D) FHM, as a function of temperature and doping (or excess charge carrier density), is depicted in \figref{fig:fig1}c, as inferred in part from recent computational studies\cite{qin:2022}.
It features an insulating ground state at half-filling with antiferromagnetic (AFM) long-range order, which becomes stripe (magnetic and charge) ordered when doped.
Superconducting long-range order has also been reported upon doping, with the delicate interplay between superconductivity and stripe ordering depending on details of the model \cite{qin:2020,xu:2024}.
The two-dimensional Fermi-Hubbard model is widely regarded as capturing essential qualitative features of strongly correlated materials such as the cuprate compounds\cite{anderson:1987,dagotto:1994}, which are notably known for exhibiting high-$T_{\mathrm{c}}$ superconductivity.

At higher temperatures, the metallic phase of the Fermi-Hubbard model features a ``pseudogap'' below a characteristic 
temperature $T^{*}$~\cite{tremblay:2006,gull:2010,wu:2018,simkovic:2024,qin:2022}, \emph{i.e.} a partial depletion of the density of states at the Fermi level~\cite{timusk:1999,norman:2005,alloul:2014}.
This pseudogap is analogous to the one found in the underdoped region of cuprates, especially when magnetic correlations are predominant, and was first identified in such materials through magnetic susceptibility measurements \cite{alloul:1989,johnston:1989}.
The physical origin of the pseudogap in cuprates has long been debated.
Some explanations emphasise its connection to magnetic correlations~\cite{alloul:2014,gunnarsson:2015,wu:2017,frachet:2020} and relation to the stripe phase~\cite{kivelson:2003,schlomer:2025}.
Possible magnetic or charge ordering instabilities of the pseudogap state have also been widely discussed~\cite{varma:1999,fauque:2006,shen:2005}.
It is sometimes also considered as a precursor of the superconducting state~\cite{norman:2005}, similar to the phenomenology found in unitary Fermi gases, where a pseudogap develops above the critical temperature for superfluidity \cite{feld:2011, gaebler:2010, li:2024a}

Over the past few years, tremendous progress has been made both from an experimental and theoretical perspective towards a better understanding of the microscopic nature of strongly correlated phases emerging in the FHM and its derivatives.
Quantum gas microscopes, in particular, have provided valuable insights into the formation of spin-ordered phases \cite{boll:2016a, mazurenko:2017a, xu:2025}, the interplay between spin and charge degrees of freedom \cite{chiu:2019a, koepsell:2019, ji:2021, prichard:2024, salomon:2019, lebrat:2024, xu:2023, hartke:2023}, the emergence of charge-ordered states \cite{hirthe:2023, bourgund:2025}, and transport properties \cite{brown:2019a, nichols:2019, guardado-sanchez:2020}.
A systematic exploration of the 2D FHM, varying both temperature and doping, is however still lacking.
Numerical methods, on the other hand, are increasingly arriving at a consensus regarding the physics of the 2D FHM in certain regimes~\cite{qin:2022,leblanc:2015,schafer:2021}.
Stripe order has been confirmed by numerous methods as the ground state at finite doping and intermediate coupling strengths \cite{leblanc:2015,zheng:2017,huang:2018a,mai:2022,xiao:2023}, the pseudogap regime has been tied to the onset of antiferromagnetic correlations \cite{gunnarsson:2015,wu:2017,wietek:2021,schafer:2021,meixner:2024,simkovic:2024} and charge clustering~\cite{sinha:2024}, and superconductivity has been found to be strongly enhanced upon adding next-nearest neighbour hopping \cite{jiang:2019,xu:2024,wietek:2022,baldelli:2025,jiang:2024}.

In this work, we perform a systematic exploration of the 2D Fermi-Hubbard model over a dense grid of dopings and temperatures (\figref{fig:fig1}c) using a quantum gas microscope Fermi-Hubbard simulator (\figref{fig:fig1}a), and uncover a novel universal scaling in spin and spin-charge correlations.
We use these correlation functions to provide a microscopic description of the underlying strongly-correlated states, especially in a regime where the presence of a pseudogap is well established\cite{gunnarsson:2015,wietek:2021,meixner:2024,simkovic:2024,wu:2017}.
We compare our measurements to different numerical methods (\figref{fig:fig1}d), including determinant Quantum Monte Carlo (dQMC) \cite{blankenbecler:1981, varney:2009}, Minimally Entangled Typical Thermal States (METTS) \cite{white:2009,stoudenmire:2010,wietek:2021,wietek:2021a} --- a method based on matrix-product states ---, and an effective model based on chargon-spinon geometric strings \cite{grusdt:2018a}.

\section{Experimental protocol}

\begin{figure*}[!thbp]
\centering
\includegraphics[scale=1]{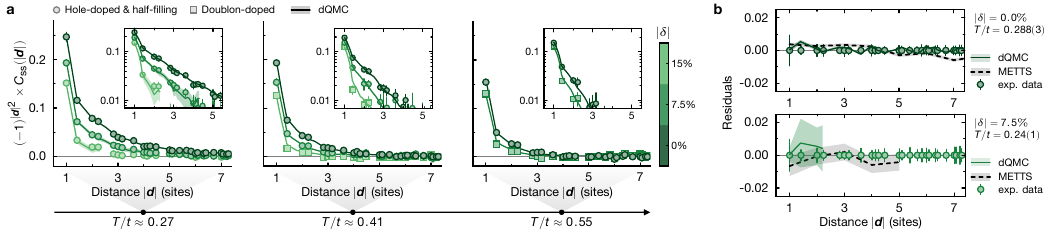}
\caption{
\textbf{Thermometry.}
\textbf{a.} The temperature of the experimental datasets is extracted by finding the best fit between dQMC numerical calculations (solid lines) and the measured spin correlations $C_{\mathrm{ss}}^{(2)}(|\bm{d}|)$ (data points), with the temperature being the only fitting parameter.
Experimental errorbars in this figure and in all figures correspond to the \SI{68}{\percent} confidence interval of a bootstrap analysis (see SI).
\textbf{b.} Residuals between dQMC and experiment (solid line) and comparison to METTS (dashed line), taking the experimental data points as a reference, for half-filling (top) and at $|\delta| = \SI{7.5}{\percent}$ doping (bottom).
Here, METTS is computed on a $32 \times 4$ cylinder, and dQMC on a square with periodic boundary conditions of size $16 \times 16$ (half-filling) or $10 \times 10$ (finite doping).
Numerical errorbars (see SI) in this figure and in all figures are indicated by shaded regions.
}
\label{fig:fig2}
\end{figure*}

Our experiment consists in loading a spin-balanced ultracold gas of $^{6}\mathrm{Li}$ atoms in a 2D optical lattice \cite{bloch:2008,esslinger:2010}, which naturally implements the FHM with Hamiltonian
\begin{equation}
	\hat H = -t\sum_{\braket{\bm{i},\bm{j}}, \sigma}\hat c^\dag_{\bm{i},\sigma}\hat c_{\bm{j},\sigma} + \mathrm{h.c.} + U\sum_{\bm i}\hat n_{\bm{i},\uparrow}\hat n_{\bm{i},\downarrow}.
\end{equation}
Here, $t$ and $U$ denote the tunneling and (on-site) interaction energies, $\sigma = \uparrow, \downarrow$ is the spin, $\hat c_{\bm{i},\sigma}$ ($\hat c_{\bm i,\sigma}^{\dag}$) is the fermionic annihilation (creation) operator for spin $\sigma$ on site $\bm i$, $\hat n_{\bm i,\sigma} = \hat c_{\bm i, \sigma}^\dag \hat c_{\bm i, \sigma}$, and $\braket{\bm i, \bm j}$ designate nearest-neighbour (NN) lattice sites.
The optical potential is engineered using a Digital Micromirror Device (DMD) such as to have a homogeneous system of $|\Omega| = \num{145}$ lattice sites surrounded by a low-density reservoir (\figref{fig:fig1}b), whose chemical potential is adjusted to control the central particle density $\braket{\hat n_{\bm{i}}}$ (SI).
The temperature of the system can be tuned by holding the atoms in the lattice for a variable time, naturally causing heating \cite{mazurenko:2017a}.
Throughout this work, tunnelling and interaction energies are set to $t/h = \SI{300(25)}{\Hz}$ and $U/h = \SI{1.95(5)}{\kHz}$, such that $U/t = \num{6.5(5)}$.
This choice is motivated by recent theoretical work showing the maximum extent of the pseudogap phase as a function of doping at temperature ranges relevant to this work~\cite{wu:2017,simkovic:2024}.
Detection is performed using a quantum gas microscope, allowing for spin and charge resolution (\figref{fig:fig1}a), as described in our recent work \cite{bourgund:2025, chalopin:2025}.

We collected \num{36485} snapshots at various dopings and heating durations and binned them accordingly.
In particular, we evaluate for each bin the system-averaged, normalised, and connected spin correlation function
\begin{equation}
	C_\mathrm{ss}^{(2)}(\bm{d}) = \frac{4}{\mathcal{N}_{\bm d}}\sum_{\bm{i} \in \Omega}\braket{\hat S_{\bm{i}}^z \hat S_{\bm{i} + \bm{d}}^z}_\mathrm{c},
\label{eq:Css}
\end{equation}
and bin together datasets with similar density and nearest-neighbour (NN) correlations $C_\mathrm{ss}^{(2)}(|\bm{d}| = 1)$, which we use as a proxy for the temperature.
In the above expression, $\mathcal{N}_{\bm d}$ is a normalisation constant that counts the number of pairs of lattice sites separated by $\bm{d}$, $\hat S_{\bm i}^z = (\hat n_{\bm i, \uparrow} - \hat n_{\bm i, \downarrow})/2$ is the $z$-component of the spin operator on lattice site $\bm i$, and $\braket{\cdot}_\mathrm{c}$ designates connected correlations.
We refer the reader to the SI for additional information regarding the exact definition of connected correlation functions, our binning procedure, and the effect of finite detection fidelity on our observables.

The above-described procedure results in about \num{120} datasets of \SI{5}{\percent}-wide doping bins, with $\SI{-30}{\percent} \leq \delta = \braket{\hat n_{\bm i}} - 1 \leq \SI{30}{\percent}$.
For each dataset, we compare the radially-averaged spin correlations $C_\mathrm{ss}^{(2)}(|\bm d|)$ to dQMC simulations performed at various dopings and temperatures (\figref{fig:fig2}a, see also the SI).
We observe excellent agreement between the experimental and numerical data with temperature as the only adjustable parameter.
This procedure thus allows the assignment of a temperature to each dataset, ranging from $T/t \approx 0.2$ to $T/t \approx 1$ (the Boltzmann constant is set to $\kB = 1$).
The results are furthermore compared to METTS (see \figref{fig:fig2}b), which also shows an excellent agreement with both the experimental and dQMC data.
The range of doping and temperature explored in this work is indicated by the hatched region in \figref{fig:fig1}c.

\section{Spin correlations}

\begin{figure*}[!thpb]
\centering
\includegraphics[scale=1]{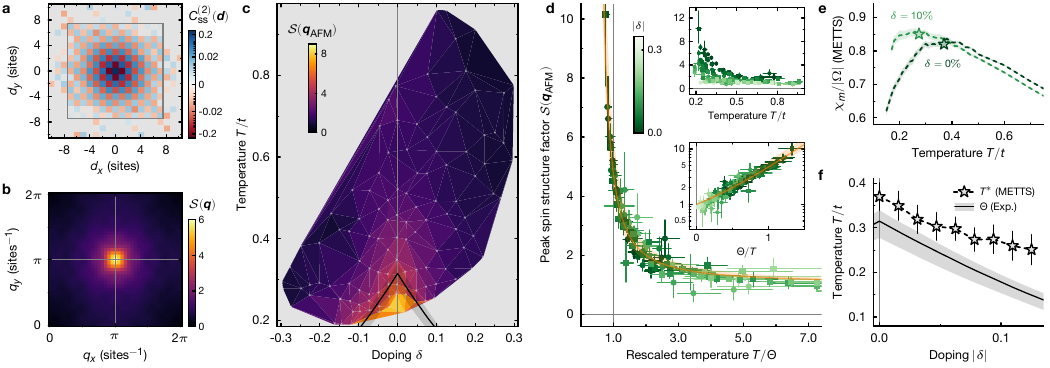}
\caption{
\textbf{Magnetic correlations in the pseudogap.}
\textbf{a.} Spin correlation map at $T/t = \num{0.25(1)}$ and $\delta \approx \SI{-2.5}{\percent}$.
\textbf{b.} Symmetrised spin structure factor $\mathcal{S}(\bm q)$ evaluated from a subregion of the spin correlation map (gray square in (\textbf{a})).
\textbf{c.} Peak structure factor $\Spp$ as a function of doping $\delta$ and temperature $T/t$.
Each vertex is a data point, and each triangle is coloured according to the average of its vertices.
The black line corresponds to the experimentally extracted doping-dependent energy scale $\scaling(\delta)$ (see panel (\textbf{d})).
\textbf{d.} Peak structure factor as a function of the rescaled temperature $T/\scaling$ (see text), with the orange line showing the exponential dependence of Eq. \eqref{eq:weak_coupling}.
Both insets show the same data, without rescaling (top) and in semi log-scale (bottom).
\textbf{e.} Magnetic susceptibility extracted from METTS numerical simulations, for $\delta = 0$ and $\delta = 0.1$.
The maximum of susceptibility (stars) occurs at $T = T^*$, is doping dependent, and marks the crossover to the pseudogap phase.
\textbf{f.} Comparison between $\scaling(\delta)$, extracted from the $\Spp$ collapse (panel (\textbf{d})) and $T^*$ extracted from METTS.
}
\label{fig:fig3}
\end{figure*}

In the thermodynamic limit, the Mermin-Wagner theorem prevents true long-range order at any finite temperature in the 2D FHM \cite{mermin:1966}.
Nevertheless, close to half-filling, AFM correlations become more and more prominent as the temperature decreases.
Here, we measure spin correlations which are significantly non-zero up to more than seven sites, a distance comparable to the radius of our system.
In order to characterise the strength and range of spin correlations, we compute the spin structure factor $\mathcal{S}(\bm q)$, defined as the Fourier transform of the spin correlations
\begin{equation}
  \mathcal{S}(\bm q) = \sum_{\bm{d} \in \Phi}\ee^{\ii \bm{qd}}C_{\mathrm{ss}}^{(2)}(\bm d),
  \label{eq:Sq}
\end{equation}
where $\Phi$, represented by the gray square in \figref{fig:fig3}a, corresponds to the region $\max(d_x, d_y) \leq \num{7}$ sites.
We show in \figref{fig:fig3}b the spin structure factor computed from the data in \figref{fig:fig3}a, depicting a strong peak centred at $\qafm = (\pi, \pi)$ as a manifestation of sizeable and extended AFM correlations.
We emphasise that, given the experimental temperature and resolution achieved here (see \figref{fig:fig1}c), we do not detect a shift of the ordering vector away from $(\pi,\pi)$ expected at low temperatures\cite{simkovic:2022, mai:2022} as a possible precursor to stripe ordering \cite{wietek:2021,xiao:2023}.
Note that the structure factor we consider here is associated with the equal-time correlation function, and not the usual static (zero-frequency) one \cite{kivelson:2003}.

We extract the peak structure factor $\Spp \equiv \mathcal{S}(\bm{q} = \qafm)$ for each dataset, and give the results in \figref{fig:fig3}c.
There, each point corresponds to a single dataset, and a Delaunay triangulation is performed to pave the $(\delta, T)$ space: the colour of each triangle is given by the average value of $\Spp$ over all three of its vertices.
As expected from the particle-hole symmetry of the Fermi-Hubbard model on a bipartite lattice \cite{arovas:2022, note:symmetry}, the peak structure factor appears symmetric with respect to half-filling, \emph{i.e.} hole doping ($\delta < 0$) and doublon doping ($\delta > 0$) have identical effects on the spin correlations.
We clearly observe an increase of $\Spp$ as dopings get closer to half-filling and the temperature is reduced.

The temperature dependence of the measured peak structure factor is found to be well described by an exponential form:
\begin{equation}
	\Spp(\delta, T) \sim \ee^{2\scaling(\delta)/T},
\label{eq:weak_coupling}
\end{equation}
with $\scaling(\delta)$ a doping-dependent temperature scale. 
This form is motivated by the relation $\Spp \sim \xi^{2}$ between the peak structure factor and the spin correlation length $\xi$, as given by the Ornstein-Zernike form \cite{schafer:2021}.
The exponential dependence for $\xi$ is indeed expected when the ground-state of a 2D quantum system displays long-range spin ordering, in which case the scale $\scaling$ identifies with the spin stiffness $2\pi\rho_s$ \cite{chakravarty:1989}.
Given the range of temperatures accessible to our experiments (see \figref{fig:fig1}c), we cannot ascertain that the exponential fit persists down to low-$T$ and thus whether the ground-state displays long-range spin order; $\scaling(\delta)$ should be viewed 
as an empirically defined spin stiffness whose doping-dependence is determined from experiments.
In essence, expression (\ref{eq:weak_coupling}) describes how a reduction in AFM order through doping can be compensated for by further cooling the system.

To determine the doping-dependence of $\scaling$, we introduce a second order expansion \emph{Ansatz} of the form $\scaling(\delta) = \scaling^{(0)} + \scaling^{(1)}|\delta| + \scaling^{(2)}|\delta|^2$, where the $\scaling^{(i)}$ coefficients are fit parameters.
The coefficient $\scaling^{(0)}$ is determined by fitting the half-filled data, to find $\scaling^{(0)}/t = \num{0.32(4)}$.
The other two coefficients are determined through an optimization procedure that aligns the $\Spp(T, \delta)$ curves to a universal form $\Spp[\scaling(\delta)/T]$, as described by \Eqref{eq:weak_coupling}.
The procedure, described in more detail in the SI, yields $\scaling^{(1)}/t = \num{-1.6(4)}$ and $\scaling^{(2)}/t = \num{2(2)}$.
The result of the optimization is shown as the black line in \figref{fig:fig3}c,f.

Remarkably, we find an excellent collapse of the magnetic peak structure factor $\Spp[\scaling(\delta)/T]$ following the above procedure for all doping and temperatures explored here (see \figref{fig:fig3}d and its lower inset).
The collapse only depends on the absolute value of doping and holds for both hole- and particle-doped systems (disk and square symbols, respectively).
In contrast, the unscaled raw data exhibits a distinctly doping-dependent behaviour (see \figref{fig:fig3}d, top inset).

We compare the extracted $\scaling(\delta)$ to the expected temperature $T^*$ that marks the crossover to the pseudogap regime in the FHM.
Although the opening of a pseudogap is nowadays commonly identified through ARPES measurements~\cite{damascelli:2003}, 
the pseudogap in cuprates was originally identified through the temperature-dependence of the magnetic susceptibility $\chi_m$\cite{alloul:1989, johnston:1989}.   
In the Hubbard model, numerical studies have revealed that the uniform susceptibility displays a maximum at a temperature $T^*$ which can serve as a definition of the pseudogap onset temperature\cite{chen:2017,wietek:2021}.
Here, the value of $T^*$ is obtained from the maximum of $\chi_m$ in METTS simulations, 
as shown in \figref{fig:fig3}e, and compared to $\scaling(\delta)$ in \figref{fig:fig3}f.
$T^*$ (from METTS) and $\scaling$ (from the experimental data) are relatively close in amplitude, and are decreasing functions of $\delta$.
The extracted value of $\scaling$ is comparable but slightly smaller than $T^*$, suggesting that the universal scaling associated with $\scaling$ that we observe experimentally is linked to the opening of the pseudogap in the system.
We stress here that we are not aware of a previous study that attempts to relate $\scaling$ --- a doping-dependent spin stiffness --- to $T^*$.
Nevertheless, Monte-Carlo simulations on the Heisenberg model have shown that the temperature $T^*$ marking the maximum of spin susceptibility coincides with the spin stiffness $2\pi\rho_s$ \cite{makivic:1991}.
Although no formal relation can be established, our results suggest that these quantities reflect the same underlying phenomenology.

The connection between the pseudogap and AFM spin-correlations is well established theoretically in the weak-coupling regime when the correlation length reaches large values.
In this regime, the Vilk-Tremblay spin fluctuation theory states that a pseudogap opens when $\xi$ exceeds the de Broglie thermal wavelength $\sim v_\mathrm{F}/\pi T$ where $v_\mathrm{F}$ is a typical Fermi velocity (\oldcite{vilk:1997}, see also \cite{schafer:2021,abanov:2003,ye:2023}). 
We note that this criterion is not satisfied here, with $1 \lesssim \xi \approx \sqrt{\Spp} \lesssim 3$, while $2 \lesssim v_\mathrm{F}/\pi T \lesssim 4$.
Nonetheless, we find that the universal scaling $T^*\sim\scaling\sim T\log\xi$ relating the pseudogap temperature to the correlation length at weak coupling appears to have a wider range of applicability, extending to the stronger coupling regime investigated here. 
Thus, our results provide evidence that the pseudogap regime in the FHM at intermediate to strong coupling is concomitant to the emergence of a strongly correlated regime marked by spin correlations of moderate spatial extension. 
This picture is indeed consistent with computational studies~\cite{gunnarsson:2015,wu:2017,wietek:2021,lihm:2025}, which also confirm that a pseudogap emerges in the range of parameters reached in our experiment~\cite{simkovic:2024}.

\section{Dopant-spin-spin correlations}
\label{sec:secIII}

\begin{figure}[!t]
\centering
\includegraphics[scale=1]{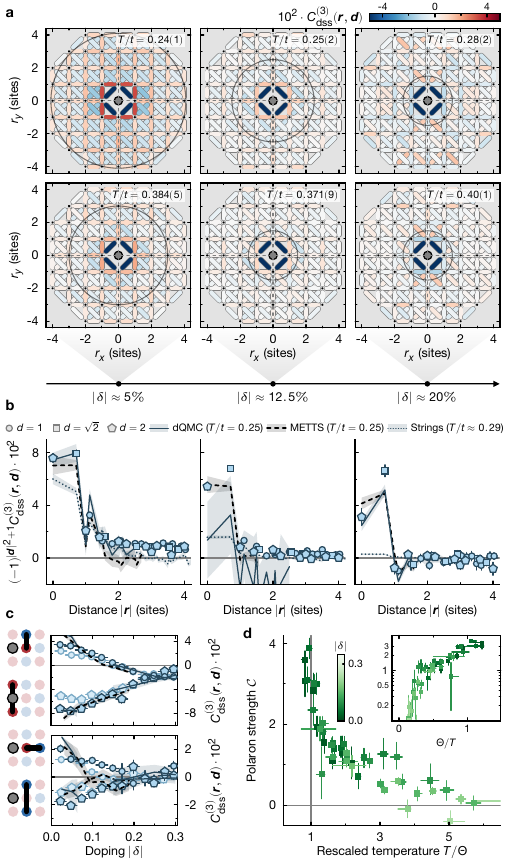}
\caption{
\textbf{Emergence of extended polarons in the pseudogap.}
\textbf{a.} Example of polaron correlations $C_{\mathrm{dss}}(\bm r, \bm d)$ at low temperature ($T/t \approx 0.25$, first row) and slightly larger temperature ($T/t \approx 0.4$, second row) and for different dopings.
The maps are symmetrised, and the circle is a guide to the eye, indicating the distance over which correlations are sizeable.
\textbf{b.} Distance-averaged, sign-corrected polaron correlations associated to the coldest datasets (first row in (\textbf{a})), for the same dopings.
Different spin bonds ($|\bm d| = 1, \sqrt{2}, 2$ for NN, NNN (diagonal) and second NN bonds, respectively) are represented by different symbols.
Solid, dashed, and dotted lines correspond to numerical simulations (dQMC, METTS, and geometric strings, respectively).
\textbf{c.} Polaronic correlations on selected bonds at short distances as a function of doping.
Data points in dark (light) blue correspond to a temperature $T/t \approx 0.25$ ($T/t \approx 0.4$).
\textbf{d.} Strength of the polaron $\mathcal{C}$ (see text) as a function of the rescaled temperature $T/\scaling$.
The inset depicts the same data in semi-log scale.
}
\label{fig:fig4}
\end{figure}

We now turn to higher-order correlations in order to explore how the interplay between dopants and spins behaves when entering the strongly correlated regime associated with $T \leq \scaling \simeq T^*$.
In particular, we evaluate third-order, connected, and normalised dopant-spin-spin correlations as
\begin{equation}
C_{\mathrm{dss}}^{(3)}(\bm r, \bm d) = \frac{4}{\mathcal{N}_{\bm{r}, \bm d}} \sum_{\bm i \in \Omega} \frac{\braket {\hat d_{\bm{i}} \hat S_{\bm i + \bm r + \bm d/2}^z \hat S_{\bm i + \bm r - \bm d/2}^z}_\mathrm{c}}{\braket{\hat d_{\bm i}}},
\label{eq:polaron}
\end{equation}
where $\mathcal{N}_{\bm r, \bm d}$ is a normalisation constant.
$C_{\mathrm{dss}}^{(3)}$ evaluates how spin correlations between two spins separated by $\bm d$ are affected by the presence of a dopant at distance $\bm r$ from the spin bond, and characterises the spatial structure of magnetic polarons \cite{koepsell:2019,prichard:2024, xu:2023}.
We limit our analysis to NN ($|\bm d| = 1$), NNN ($|\bm d| = \sqrt{2}$) and second NN ($|\bm d| = 2$) spin bonds.
For the remainder of the manuscript, we rely on the particle-hole symmetry of the Hubbard model and combine both hole- and doublon-doped sectors to improve statistics \cite{note:symmetry} (see also SI).
Thus, in \Eqref{eq:polaron}, the operator $\hat d_{\bm i}$ is the dopant operator at lattice site $\bm i$ (SI).
The experimental measurements of higher-order correlators are compared to calculations from METTS, dQMC and from the geometric string model.
In the case of METTS and the string model, classical snapshots akin to that of the quantum gas microscope are sampled (SI).

In \figref{fig:fig4}a, we present examples of symmetrised polaron correlation maps at different dopings and temperatures for nearest-neighbour (NN, $|\bm{d}| = 1$) and NNN ($|\bm{d}| = \sqrt{2}$) spin bonds.
Close to half-filling ($|\delta| \approx \SI{5}{\percent}$) and at low temperatures, we observe significant non-zero correlations over a disk of radius $|\bm r| \approx 4$ around the dopant (grey circles in \figref{fig:fig4}a), indicating that a single dopant affects the background AFM order over a surrounding region that spans more than \num{50} lattice sites.
As either the doping or the temperature increases, the size of the polaron decreases.

In \figref{fig:fig4}b, we show the sign-corrected and distance-averaged correlations to clearly evaluate the range over which connected correlations remain non-zero for the coldest datasets ($T/t < 0.3$).
We systematically observe non-zero correlations at short ranges --- \emph{i.e.} in the core of the magnetic polarons corresponding to the spin bonds in the direct vicinity of the dopant --- for all temperature and doping values studied here (hatched region in \figref{fig:fig1}c), with a larger amplitude for colder datasets (see SI for a temperature comparison).
However, close to half-filling, we discover the emergence of longer-range correlations marked by a non-zero tail in the sign-corrected correlations.
Such a low-temperature feature was not observed in previous studies pertaining to the FHM on a square lattice \cite{koepsell:2019,koepsell:2021a,hartke:2023}.
The core structure of the polarons are qualitatively well reproduced by dQMC and METTS simulations (solid and dashed lines in \figref{fig:fig4}).
While the low-doping tail structure is not exhibited by METTS on the $32 \times 4$ cylinder, the geometric string model (dotted lines) --- calculated using the coldest undoped dataset ($T/t \approx 0.29$, see SI) --- shows a larger spatial extension despite a reduced core amplitude, and the dQMC is compatible with the onset of a tail.
At larger doping, the geometric string calculations fail to reproduce the experimental data; we attribute this discrepancy to the relatively large contribution of doublon-hole fluctuations in our parameter regime ($U/t \approx 6.5$, see also SI).
Note that the dQMC calculations around \SI{12.5}{\percent} are strongly affected by the sign problem \cite{tarat:2022}, resulting in very large uncertainties.

At larger dopings ($|\delta| \gtrsim \SI{20}{\percent}$), the NN spin bond closest to the dopant changes sign.
This feature, as well as a much-reduced temperature dependence, can be attributed to Fermi-Liquid behaviour, in which the hole and spin correlations are dominated by the Pauli principle \cite{cheuk:2016a, koepsell:2021a}.
Such behaviour sharply contrasts with the low doping behaviour, where one sees the emergence of extended polarons at a temperature consistent with $T \lesssim \scaling$.
These observations are illustrated in more detail in \figref{fig:fig4}c, where $C_{\mathrm{dss}}^{(3)}(\bm r, \bm d)$ is given as a function of doping for a few bonds closest to the dopant.

In analogy to the magnetic structure factor, which quantifies AFM strength, we propose to introduce a ``polaron strength'' $\mathcal{C}$ by integrating the sign-corrected correlations, $\mathcal{C} = \sum_{\bm r, \bm d}(-1)^{|\bm d|^2 + 1}C_{\mathrm{dss}}(\bm r, \bm d)$.
In particular, we expect that the emergence of a long-range polaronic tail contributes significantly to $\mathcal C$.
We show in \figref{fig:fig4}d the polaron strength as a function of the rescaled temperature $T/\scaling$.
In the range of temperatures and dopings explored here (hatched region in \figref{fig:fig1}c), we notice that the polaron strength follows a similar universal trend as found for the spin ordering.
Specifically, it increases significantly when $T \leq \scaling$, signalling the importance of polaronic spin-charge correlations in the pseudogap regime.
In the intermediate regime $1 \lesssim T/\scaling \lesssim 2$, $\mathcal{C}$ is dominated by the polaronic core, whose amplitude decreases as $T/\scaling$ increases.
At large dopings ($T \gg \scaling$ in our datasets) the presence of sign-changed spin bonds cancels out the contribution from the polaronic core, such that $\mathcal{C}$ goes to zero.

\section{Higher-order correlations}
\label{sec:secIV}

\begin{figure}[!t]
\centering
\includegraphics[scale=1]{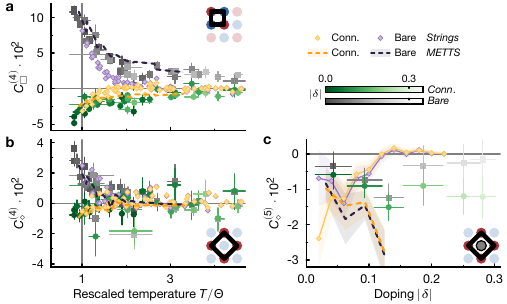}
\caption{
\textbf{Higher-order correlations.}
\textbf{a, b} $4^\mathrm{th}$-order spin correlations with four spins arranged in a square (\textbf{a}), 
or in a diamond shape (\textbf{b}).
The connected (bare) correlations, represented by green circles (grey squares), are plotted against the rescaled temperature $T/\scaling$.
The colour scale indicates the doping.
\textbf{c.} $5^\mathrm{th}$-order spin-charge correlations 
in the diamond configuration, showing that the presence of a dopant strongly affects the correlations. In all panels, the dashed lines and the dots correspond to METTS and geometric string calculations, respectively.
}
\label{fig:fig5}
\end{figure}

Quite generally, non-zero correlations beyond second order constitute key signatures of strongly correlated regimes, as weakly correlated systems are well characterised by effective quasi-particle theories with Gaussian --- and thus low-order --- correlations.
The third-order polaronic correlations explored above confirm the existence of a strongly correlated regime for $T \leq \scaling \simeq T^*$, associated with the pseudogap phase of the FHM.
A natural extension consists in exploring higher-order correlations, which are accessible with our experimental apparatus.

We begin by considering the $4^{\mathrm{th}}$-order normalised and connected spin correlations, defined as
\begin{equation}
	C_{\mathcal{K}}^{(4)} = \frac{2^{4}}{\mathcal{N}_\mathcal{K}}\smashoperator{\sum_{\substack{\bm i \in \Omega \\ \bm i, \bm j, \bm k, \bm \ell \in \mathcal{K}}}}\braket{\hat S_{\bm i}^z \hat S_{\bm j}^z \hat S_{\bm k}^z \hat S_{\bm \ell}^z}_{\mathrm{c}},
\label{eq:4point}
\end{equation}
where $\mathcal{K}$ designates a spatial arrangement of the four spins, and $\mathcal{N}_\mathcal{K}$ denotes the corresponding normalisation factor.
Here, we consider NN bonds or diagonal bonds only, \emph{i.e.} spins arranged in a square ($C_\Box^{(4)}$), 
or in a diamond-shape ($C_\diamond^{(4)}$) configuration (see SI for additional data corresponding to a T-shape configuration).
The results are shown in \figref{fig:fig5}a,b, where both the bare and the connected correlations are plotted against the rescaled temperature $T/\scaling$.
As expected from AFM order, the bare correlations are positive
and increase in amplitude when the temperature is reduced.
In the square 
configuration, the connected correlations become significantly non-zero and negative 
as $T \lesssim \scaling$, and corroborate the emergence of a strongly correlated regime with sizeable quantum fluctuations stemming from the underlying quantum antiferromagnet.
In the case of the diamond-shaped configuration, which corresponds to more distant spins, connected correlations remain close to zero.
Overall, our results suggest that the role of quantum fluctuations, central in the emergence of strongly correlated states and non-zero high-order correlations, weakens as the distances between spins increase.
In all cases, METTS (dashed lines) and geometric string (diamonds) calculations agree well with the experimental data.

Interestingly, we find significant qualitative and quantitative changes in the diamond-shape configuration when evaluated \emph{conditioned} on the presence of a dopant at its centre.
More precisely, we evaluated the 
$5^{\mathrm{th}}$-order normalised conditional dopant-spin correlations \cite{bohrdt:2021b} (see SI for details),
\begin{equation}
	C_{\mathcal{K}}^{(5)} = \frac{2^{4}}{\mathcal{N}_\mathcal{K}}\smashoperator{\sum_{\substack{\bm i \in \Omega \\ \bm i, \bm j, \bm k, \bm \ell, \bm m \in \mathcal{K}}}}\left.\frac{\braket{\hat d_{\bm i} \hat S_{\bm j}^z \hat S_{\bm k}^z \hat S_{\bm \ell}^z \hat S_{\bm m}^z}_{\mathrm{c}}}{\braket{\hat d_{\bm i}}}\right|_{\mathrm{cond.}\ \hat{d}_{\bm i}}.
\label{eq:5point}
\end{equation}
To compensate for more limited statistics, the subsequent analysis employs broader doping intervals (\SI{10}{\percent} rather than the \SI{5}{\percent} used previously).
In \figref{fig:fig5}c, 
$C_\diamond^{(5)}$ is plotted against doping for $T/t \approx 0.25$ (see SI for the T-shape configuration and for a comparison with higher temperatures).
Close to half-filling, both the bare and conditioned correlations vanish due to dominant effects of doublon-hole fluctuations --- each hole (doublon) is in the direct vicinity of a doublon (hole) and thereby voids the correlations
However, we measure significant non-zero values for $C_\diamond^{(5)}$ around $|\delta| = \num{10}-\SI{20}{\percent}$ and, more importantly, the values are equal in amplitude to the bare correlations.
This is in stark contrast to $C_\diamond^{(4)}$ (which is governed by the bare correlator) and suggests that the diamond-shape correlator is dominated by higher-order terms in the presence of a dopant.

Such a feature --- dominant higher-order correlation --- is characteristic of strongly correlated systems and can here be attributed to the presence of fluctuating chargon-spinon structures.
It indicates that, upon entering the pseudogap, magnetic structures emerge that can only be revealed through higher-order correlations (and hence lie beyond our conventional understanding of Gaussian fluctuations).
Our results furthermore suggest that these string-like correlations in the vicinity of dopants persist up to about \SI{20}{\percent} doping, providing an intuitive microscopic understanding of the nature of moving dopants and the formation of geometric strings \cite{bohrdt:2021b}.
The METTS calculations (dashed lines), converged up to \SI{12.5}{\percent} doping, exhibit a trend consistent with the experimental data.
The string-based calculation approach also captures the overall features of the experimental data, notably reproducing the dominance of higher-order correlations at finite doping.
However, the string calculations show significant correlations only up to about \SI{10}{\percent} doping.
This disagrees notably with METTS and with the experimental data, similarly to the dopant-spin-spin correlations (\figref{fig:fig4}b).
This discrepancy indicates that the geometric string framework --- intrinsically linked to the physics of the weakly doped antiferromagnetic parent compound --- does not quantitatively reproduce dopant-spin correlations within our parameter regime, although it remains accurate in capturing magnetic correlations (\figref{fig:fig5}a,b).

\section{Discussion and Outlook}

In this work, we have explored the Fermi-Hubbard model with a quantum gas microscope in a regime of temperature and doping in which numerical studies predict the existence of a pseudogap (see \figref{fig:fig1}c) \cite{wu:2017,simkovic:2024,wietek:2021}.
Our results, based on the evaluation of spin and dopant-spin correlations, reveal the emergence of a strongly correlated regime concomitant to the pseudogap.
The peak spin structure factor, in particular, allows to extract a doping-dependent temperature scale $\scaling$, which marks the transition to the strongly correlated regime.
The temperature scale not only organises the spin correlations into a universal scaling form but also matches the pseudogap temperature $T^*$ extracted independently from the saturation of the magnetic susceptibility using METTS (\figref{fig:fig3}).
Recent unpublished numerical studies furthermore suggest that the energy scale $\scaling$ reproduces well the doping-dependence of low-energy magnon excitations, thus providing an additional link to the underlying magnetic origin of the pseudogap \cite{demler:discussion}.
Taken together, these two findings imply that the pseudogap emerges concurrently with enhanced, albeit not infinitely extended, AFM correlations.
Thus, $\scaling$ serves as a proxy for the underlying spin stiffness-like scale that sets the stage for pseudogap formation.
Our work furthermore shows that this temperature scale governs the scaling behaviour of higher-order spin and spin-charge correlations.
In the temperature regime close to $T \sim \scaling$, we observe extended third-order polaronic dopant-spin correlations beyond the direct vicinity of the dopant (\figref{fig:fig4}).
Moreover, we have measured fifth-order dopant-spin correlations for which the contribution of lower-order terms is limited, suggesting that genuine quantum many-body correlations --- beyond two or even three particles --- dominate the system's physical properties upon entering the pseudogap.
These higher-order correlators can additionally serve as benchmarks for testing the validity of different theoretical approaches.

Our work opens numerous perspectives for a better experimental understanding of the pseudogap phase in the Fermi-Hubbard model, with our main conclusions going in favour of a strong relation between spin correlations and the pseudogap.
It was also suggested \cite{emery:1995, mishra:2014} that the pseudogap phase, in analogy to the pseudogap in the context of the BEC-BCS crossover of strongly-correlated Fermi gases, corresponds to the emergence of preformed pairs as a precursor to a superconducting state.
Future studies beyond the scope of this work include the exploration of pairing among dopants, especially in the regime of extended polaronic tails (\figref{fig:fig4}), where significant dopant-spin correlations are measured at distances comparable to the mean distance between dopants $|\delta|^{-1/2}$.
The emergence of hidden order has also been proposed as a characteristic of the pseudogap \cite{varma:1999,fauque:2006,chakravarty:2001,schlomer:2025}, and could be explored using high-order correlation functions, which by nature reveal correlations beyond the two-point (Gaussian) level.
At lower temperatures, one furthermore expects the emergence of a charge density wave (stripes), also associated with a shift of the peak in the spin structure factor away from $(\pi,\pi)$.
Such signatures were not observed in our experimental data and likely require reaching colder temperatures. 
This would also allow one to probe the possible connection of the pseudogap with the softening of the charge response associated with a maximum of the compressibility (Widom line~\cite{sordi:2012}).

The momentum-space structure of the pseudogap is also an outstanding question for future studies.
In the solid-state context, momentum-resolved measurements such as angular-resolved photoemission spectroscopy (ARPES)~\cite{damascelli:2003, sobota:2021} have revealed that the suppression of the density of states associated with the pseudogap occurs near the anti-nodal points of the Brillouin zone.
Momentum-resolved measurements of the single-particle spectral function have so far remained out of reach of quantum gas microscope experiments, except in very specific configurations (see \emph{e.g.} \cite{brown:2020}, in the case of the attractive Hubbard model).
Nevertheless, ARPES techniques adapted to optical lattices~\cite{dao:2007} and quantum gas microscopy~\cite{bohrdt:2018} have been proposed, offering a promising future direction. These approaches could enable the simultaneous acquisition of spectroscopic information and microscopic real-space correlation data, as presented here. Moreover, they would provide a platform to test theoretical predictions concerning the interplay between Fermi surface topology and the pseudogap~\cite{wu:2018}.

~\\
\textbf{Acknowledgements:}
We thank Thomas Schäfer, Andr\'e-Marie Tremblay, Yury Vilk and Eugene Demler for insightful discussions, 
as well as David Clément and Sarah Hirthe for careful reading of the manuscript.
This work was supported by the Max Planck Society (MPG), the Horizon Europe program HORIZON-CL4-2022 QUANTUM-02-SGA (project 101113690, PASQuans2.1), the German Ministry of Education and Research (BMBF grant agreement 13N15890, FermiQP), and Germany's Excellence Strategy (EXC-2111-390814868). A.W.\ acknowledges support by the DFG through the Emmy Noether program (Grant No.\ 509755282). The Flatiron Institute is a division of the Simons Foundation. This research was supported in part by grant NSF PHY-2309135 to the Kavli Institute for Theoretical Physics (KITP). F.G. acknowledges support from the European Research Council (ERC) under the European Union's Horizon 2020 research and innovation programme (Grant Agreement no 948141) — ERC Starting Grant SimUcQuam.

~\\
\textbf{Author contributions:}
T.C. led the project.
T.C and P.B. contributed significantly to data collection.
T.C. analysed the data.
A.S. and A.W. performed the METTS simulations and the associated data analysis.
L.P. and Z.W. performed the dQMC calculations.
A.B. and F.G. performed the geometric string calculations.
T.C. and A.G. wrote the manuscript.
I.B. and T.H. supervised the study.
All authors worked on the interpretation of data and contributed to the final manuscript.

~\\
\textbf{Competing interests:}
The authors declare no competing interests.

~\\
\textbf{Data availability}
The datasets generated and analysed during the present study, as well as the code used for the analysis, are available from the corresponding author upon reasonable request.

\bibliographystyle{naturemag}

\clearpage
\newpage
\makeatletter
\renewcommand{\thefigure}{S\arabic{figure}}
\renewcommand{\theequation}{S\arabic{equation}}

\renewcommand{\theHfigure}{S\arabic{figure}}
\renewcommand{\theHequation}{S\arabic{equation}}
\makeatother
\setcounter{figure}{0}
\setcounter{table}{0}
\setcounter{section}{0}
\setcounter{equation}{0}
\appendix

\makeatletter
\let\oldtitle\@title
\title{Supplementary information for: \\ \oldtitle}
\makeatother

\maketitle

\subsection{System preparation}

Our quantum simulator realises the spin-1/2 repulsive Fermi-Hubbard model using ultracold $^{6}$Li atoms in 2D optical lattices with lattice constants $a_x = \SI{1.11(1)}{\micro\m}$ and $a_y = \SI{1.15(1)}{\micro\m}$ generated from the interference of \SI{532}{\nm} laser beams under an angle of about \SI{27}{\degree}.
Additional technical details can be found in our recent work \cite{chalopin:2025,bourgund:2025} and in references therein.
The smallest temperatures reported in this work correspond to a sequence where a degenerate Fermi gas ($T/T_\mathrm{F} < 0.1$, with $T_\mathrm{F}$ the Fermi temperature) is loaded in the optical lattices with a \SI{200}{\ms} quadratic ramp.
We use an additional holding time in the lattices up to \SI{700}{\ms} to cause heating up to $T/t \approx 1$.

The NN tunneling $t$ is set by the lattice depth ($V_x = \num{6.5(3)}E_{\mathrm{R}, x}$ and $V_y = \num{6.3(3)}E_{\mathrm{R}, y}$, where $E_{\mathrm{R}, i} = h^2/8Ma_i^2$, $i = x, y$ is the lattice recoil energy), and the onsite interaction energy $U$ is adjusted using the broad Feshbach resonance of $^{6}$Li around \SI{830}{\Gauss}.
At such lattice depths, the second NN hopping amplitude is about \SI{3.5}{\percent} of the NN hopping amplitude ; it is neglected throughout this work, but could introduce systematic discrepancies between the experiment and numerics.
The NNN diagonal coupling, however, vanishes due to the separability of our square lattice.

As mentioned in the main text (see \figref{fig:fig1}b), we employ tailored Digital-Micromirror-Device (DMD) potentials to engineer a physical system ($\Omega$) surrounded by a low-density reservoir.
The energy difference $\mu_{\rm{r}}$ between the system and the reservoir acts as a chemical potential, and is tuned close to half the interaction energy $U/2$ for data at half-filling $\braket{\hat n_i} = 1$.
For doublon (hole) doped data, $\mu_{\rm r}$ is increased (decreased).

\subsection{Finite fidelity corrections}
\label{sec:correction}

\begin{figure}[!t] \centering
\includegraphics[scale=1]{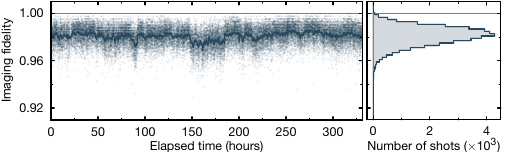}
\caption{
\textbf{Evaluation of finite imaging fidelity.}
Time trace of the imaging fidelity $\mathcal{F}$ per snapshot (blue dots) and 2h running average (solid line), used for fidelity correction.
The histogram shows the imaging fidelity over the whole dataset, yielding $\mathcal{F} = \num{0.981(8)}$.
}
\label{fig:figS1}
\end{figure}

Our apparatus is designed to perform spin-resolved imaging by physically separating both spin states $\sigma = \uparrow, \downarrow$ in two different imaging planes which are individually imaged \cite{koepsell:2020}.
At each experimental run, two subsequent fluorescence images of each imaging plane are acquired with \SI{1.5}{\s} exposure time each, separated by \SI{1.5}{\s} of ``dark'' time during which the atoms fluoresce while the camera shutter is closed.
The imaging fidelity for each plane is evaluated by comparing both images,
\begin{equation*}
    \mathcal{F}_{\sigma} = \left[\frac{\sum_{\bm i \in \Omega}{n_{\bm i,\sigma}^{(1)}n_{\bm i,\sigma}^{(2)}}}{\sum_{\bm i\in\Omega}n_{\bm i,\sigma}^{(1)}}\right]^{1/3}.
\end{equation*}
Here, $n_{\bm i, \sigma}^{(\alpha)}$ is the \emph{detected} occupancy of spin $\sigma$ atoms at site $\bm i$ on the first ($\alpha = 1$) or second ($\alpha = 2$) picture.
The average fidelity is $\mathcal{F} = (\mathcal{F}_\uparrow + \mathcal{F}_\downarrow)/2$.
A rolling average over a period of 2h is then performed on $\mathcal{F}$, which is then used for evaluating fidelity-corrected correlation functions (\figref{fig:figS1}).

We follow an analysis similar to the one performed in \cite{koepsell:2021a}.
We introduce hole ($\hat h_{\bm i}$), singlon ($\hat s_{\bm i}$), doublon ($\hat \eta_{\bm i}$) and spin ($\hat S_{\bm i}^{z}$) operators from local density operators as
\begin{equation*}
\begin{aligned}
    \hat h_{\bm i} & = (1-\hat n_{\bm i, \uparrow})(1-\hat n_{\bm i, \downarrow}) \\
    \hat s_{\bm i} & = \hat n_{\bm i, \uparrow} + \hat n_{\bm i, \downarrow} - 2\hat n_{\bm i, \uparrow}\hat n_{\bm i, \downarrow} \\
    \hat \eta_{\bm i} & = \hat n_{\bm i, \uparrow}\hat n_{\bm i, \downarrow} \\
    \hat S_{\bm i}^{z} & = (\hat n_{\bm i, \uparrow} - \hat n_{\bm i, \downarrow})/2.
\end{aligned}
\end{equation*}
Assuming that the finite fidelity $\mathcal{F} = 1-p$ is only due to particle losses, we introduce \emph{detected} operators as (the site index $\bm i$ is omitted for clarity)
\begin{equation*}
\begin{aligned}
	\hat h_\mathrm{det} & = \hat h + p\hat s + p^2\hat \eta \\
	\hat s_\mathrm{det} & = (1-p)\hat s + 2p(1-p)\hat \eta \\
	\hat \eta_\mathrm{det} & = (1-p)^2\hat \eta \\
	\hat S^{z}_{\mathrm{det}} & = (1-p)\hat S^{z}.
\end{aligned}
\end{equation*}
We emphasise that these detected operators are the ones directly measured by the experimental apparatus.
These relations can be inverted to recover the ``true'' operators,
\begin{equation*}
\begin{aligned}
	\hat \eta & = (1-p)^{-2}\hat \eta_\mathrm{det} \\
	\hat s & = (1-p)^{-1}\hat s_\mathrm{det} - 2p (1-p)^{-2}\hat \eta_\mathrm{det} \\
	\hat h & = \hat h_\mathrm{det} - p(1-p)^{-1}\hat s_\mathrm{det} + p^2(1-p)^{-2}\hat \eta_\mathrm{det} \\
	\hat S^{z} & = (1-p)^{-1}\hat S^{z}_{\mathrm{det}}.
\end{aligned}
\end{equation*}
All correlation functions presented in this work are evaluated from the corrected operators $\hat h$, $\hat s$, $\hat \eta$ and $\hat S^z$.

\subsection{Binning procedure}

We associate to each snapshot a doping level $\delta$ given by
\begin{equation*}
    \delta = \frac 1 {|\Omega|}\sum_{\bm i \in \Omega}(\hat s_{\bm i} + 2\hat \eta_{\bm i}) - 1,
\end{equation*}
where $|\Omega|$ is the system's volume.
Hole-doped (doublon-doped) systems have $\delta < 0$ ($\delta > 0$).

\begin{figure}[!t]
\centering
\includegraphics[scale=1]{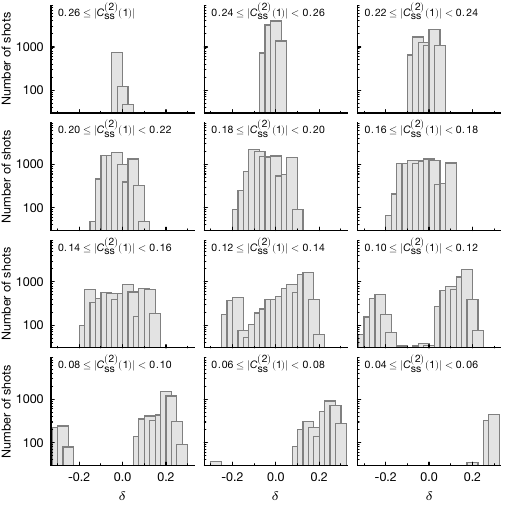}
\caption{
\textbf{Data binning.}
Number of experimental snapshots as a function of the doping bin, with the NN spin correlation bin indicated in each panel.
The vertical axis is in log scale, and the width of the (overlapping) bars represent the doping bin widths.
}
\label{fig:figS3}
\end{figure}

The data is then sorted in analysis bins:
\begin{itemize}
    \item ``Preparation'' bins, corresponding to data with identical preparation sequences --- the holding time which induces heating.
    \item Overlapping doping bins, of width \SI{5}{\percent} and separated by \SI{5}{\percent} (each shot belongs to two doping bins at most).
    \item Overlapping time bins of width 16h and separated by 16h (each shot belongs to two time bins at most).
\end{itemize}
Only bins containing at least 30 snapshots are kept.
As mentioned in the main text, we evaluate the NN spin correlations $C_\mathrm{ss}^{(2)}(|\bm d| = 1)$ on each bin as a proxy for temperature.
The data is thus binned according to $C_\mathrm{ss}^{(2)}(|\bm d| = 1)$ (width \num{2e-2}, no overlap) --- in other words, snapshots belonging to separate time bins but with similar spin-correlations are grouped.
We show in \figref{fig:figS3} the number of shot per data bin.

\subsection{Combining hole- and doublon-doped data}

The plain vanilla Fermi-Hubbard model on a bipartite lattice is particle-hole symmetric --- \emph{i.e.} invariant, up to a constant energy, under the transformation $\hat c_{\bm i, \sigma}^{\dag} \to \gamma_{\bm i} c_{\bm i, \sigma}$, where $\gamma_{\bm i} = \pm 1$ on different sublattices --- meaning that the physics associated to hole-doped and particle-doped systems is identical.
This is well verified for instance by the symmetry of \figref{fig:fig3}c.
We make use of this property to improve our statistics by mixing both sectors when evaluating higher-order observables.
In particular, we define a dopant operator at site $\bm i$, $\hat d_{\bm i}$, as
\begin{equation*}
    \hat d_{\bm i} = \left\{
    \begin{aligned}
        \hat h_{\bm i} & \quad \text{if $\delta < 0$} \\
        \hat \eta_{\bm i} & \quad \text{if $\delta > 0$.}
    \end{aligned}\right.
\end{equation*}
Note that the fidelity correction described in \secref{sec:correction} still applies.

\subsection{Bootstrap procedure and error bar estimation}

Experimental errorbars are estimated using a bootstrapping procedure (see \emph{e.g.} \cite{efron:1986}).
Each dataset is resampled 500 times, with replacement, such that each resampled dataset has the same size as the original.
Each observable --- including temperature --- is evaluated for all 500 pseudosamples.
The resulting histograms of values typically exhibit a bell-shaped distribution, from which we extract the median, as well as the $\num{16}^{\rm th}$ and $\num{84}^{\rm th}$ percentiles.
These percentiles define a \SI{68}{\percent} confidence interval and represent the statistical uncertainty.
We use them as the error bars on our experimental data points.
Note that this method yields asymmetric error bars; for values quoted in the main text, we use the standard deviation as a statistical uncertainty.

In the specific case of temperature, we furthermore take into account the finite error of the dQMC simulations (see next section below), that is propagated in the fitting procedure.
The total error bar is then taken as the root of the quadratic sum of both errorbars.

\subsection{Determinant quantum Monte Carlo}

Determinant quantum Monte Carlo (dQMC) is a stochastic method for interacting fermionic systems at finite temperature based on a path-integral representation of the partition function, in which the weights of the paths can be expressed as determinants~\cite{blankenbecler:1981,scalapino:1981,gubernatis:2016}.
The path integral representation arises for the partition function $Z$ as follows,
\begin{equation}
Z = {\rm Tr} \, \ee^{- \beta \hat{K}}
{} = \lim_{N \to \infty} {\rm Tr} \,  \left[ \exp(- \Delta \tau \hat{K}) \right]^N,
\end{equation}
where we introduced the Trotter time step $\Delta \tau = \beta/N$. The method is formulated in the grand-canonical ensemble; hence, the Hamiltonian is shifted as $\hat{K} = \hat{H} - \mu \hat{N}$.
For the Hubbard model, one uses the occupation number basis,
\begin{eqnarray}
\ket{\psi_{\Omega}} &= & \ket{\psi_{\Omega_{\uparrow}}} \ket{\psi_{\Omega_{\downarrow}}} \nonumber \\
{} & = & \ket{n_{1\uparrow}, n_{2\uparrow}, \ldots, n_{\Omega\uparrow}} \ket{n_{1\downarrow}, n_{2\downarrow}, \ldots, n_{\Omega\downarrow}}.
\end{eqnarray}
Single-particle operators, like the kinetic energy operator $\hat T = -t \sum_{\langle \bm{i}, \bm{j} \rangle, \sigma} \hat{c}^{\dagger}_{\bm{i},\sigma} \hat{c}_{\bm{j}, \sigma} + {\rm h.c.}$, transform such a basis state into another basis state (or into the same  basis state in case of the number operator $\hat{N}$).
In order to deal with two-particle operators, such as the Hubbard interaction  $U (\hat{n}_{\uparrow} - 1/2)(n_{\downarrow} - 1/2)$, one replaces in the primitive approximation
\begin{equation}
    \ee^{- \Delta \tau \hat{K}} \approx \ee^{- \Delta \tau (\hat{T} - \mu \hat{N})} e^{-\Delta \tau U \left( \hat{n}_{\uparrow} - 1/2\right) \left( \hat{n}_{\downarrow} - 1/2\right) },
\end{equation}
the exponential of the two-body operator into a sum over exponentials of one-body operators coupled to a Hubbard-Stratonovich (HS) field.
For instance, for $U >0$,
\begin{multline}
\ee^{\Delta \tau \frac{U}{2} \left( \hat{n}_{\uparrow} - \hat{n}_{\downarrow} \right)^2} = \frac{1}{4}\sum_{x = \pm 1, \pm 2} \gamma(x)\ee^{\sqrt{\frac{1}{2}\Delta\tau U} \eta(x)(\hat{n}_{\uparrow}-\hat{n}_{\downarrow})} \\ + \mathcal{O}(\Delta\tau U)^4,
\label{eq:HS} 
\end{multline}
where $ \gamma(x) $ and $ \eta(x)$ are documented in Ref.~\cite{assaad:1997, assaad:2022a},  
and the auxiliary field is seen to couple to the spin~\cite{hirsch:1983}.
In a stochastic process, this means that one generates, with equal probability, per time slice and per lattice site a HS field $x =\pm 1$ and $\pm 2$.
We note that different types of HS decompositions exist and which one is used depends on the symmetries, parameters, and quantities of interest.
For the doped Hubbard model the decomposition in the spin channel as in Eq.~\ref{eq:HS} is used, but at half filling a decoupling in the charge channel is advantageous~\cite{rombouts:1998,rombouts:1999,assaad:1999,batrouni:1990a}.

A complete set of basis states is inserted between any two exponentials of single-particle operators,
\begin{eqnarray}
Z & \approx & \sum_{\{ \psi \}} \sum_{ \{ \bm{x} \} } \braket{\psi_{\Omega} | \ee^{- \Delta \tau \hat{T}} | \psi_{\Omega}^{(1)} } \braket{\psi_{\Omega}^{(1)} | \ee^{- \Delta \tau \hat{V}(x^{(1)})} | \psi_{\Omega}^{(2)}} \cdots \nonumber \\
{} & {} & \braket{\psi_{\Omega}^{(2N-1)} | \ee^{- \Delta \tau \hat{V}(x^{(n)})} | \psi_{\Omega}}.
\end{eqnarray}
Here, $\hat{V}(x^{(j)}), j = 1, \ldots, N$ is a shorthand notation for $ \prod_{k=1}^{\Omega} \gamma( x_k^{(j)} )\ee^{\sqrt{\frac{1}{2}\Delta\tau U}\eta(x_k^{(j)} )(\hat{n}_{k \uparrow}-\hat{n}_{k \downarrow})} $
at time step $j$.
The usefulness of this decomposition lies in Thouless' theorem, stating that the exponential operator of a one-body operator acting on a slater determinant results in a new slater determinant.
Furthermore, the matrix representation of the new Slater determinant is easily found by multiplying the matrix representation of the exponential of the one-body operator with the matrix representation of the old Slater determinant.
After tracing out the fermionic degrees of freedom, the partition function is then the sum over all weights specified by the HS configurations,
\begin{equation}
Z = \sum_{ \{ \bm{x} \} } w_{\bm{x}}^{\uparrow} w_{\bm{x}}^{\downarrow},
\end{equation}
where the weights $w_{\bm{x}}^{\sigma}$ can be expressed as determinants,
\begin{equation}
    w_{\bm{x}}^{\sigma} = \det\left[ I + B^{\sigma}(\beta,0) \right], \quad \sigma = \uparrow,\downarrow.
\end{equation}
The $B$-matrices express the time-evolution as a string of matrix products,
\begin{equation}
    B^{\sigma}(\beta,0) = B^{\sigma}(\beta,\beta - \Delta \tau) \cdots B^{\sigma}(2\Delta \tau, \Delta \tau) B^{\sigma}(\Delta \tau, 0),
\end{equation}
and are given by $B^{\sigma}(\tau,\tau - \Delta \tau) = \ee^{\Delta \tau \mu} A^{\sigma}(\tau) \ee^{\Delta \tau \hat T}$.
The A-matrices are the matrix representations of the exponentials of the HS terms, $A^{\sigma}(\tau = j \Delta \tau) = {\rm Diag} [\ee^{\sigma \gamma x_1^{(j)}}, \ee^{\sigma \gamma x_2^{(j)}}, \ldots, \ee^{\sigma \gamma x_{\Omega}^{(j)}}] $, while the exponential of the kinetic energy operator has a matrix representation $\ee^{\Delta \tau \hat T} = V \, {\rm Diag} [\ee^{\Delta \tau \lambda_1}, \ee^{\Delta \tau \lambda_2}, \ldots, \ee^{\Delta \tau \lambda_{\Omega}}] \, V^\top$, found by diagonalizing the matrix representation of $\hat T$ in the number occupation basis yielding the eigenvalues $\lambda_j, j = 1, \ldots, \Omega$ and the corresponding orhthonormal eigenvectors building the columns of the matrix $V$.
Single-particle operators like the single-particle Green function are also straightforward to compute using matrix computations.
Multi-particle observables can be computed after applying Wick's theorem.
In practice, care must be taken of a few technicalities, such as matrix stabilization techniques, and sometimes faster updating mechanisms are possible.
The algorithm is sign-free only at half filling.
Away from half filling, the sign problem limits calculations to temperatures down to $T \sim t/4$ at low doping. Snapshots of Fock states with equal, positive statistical weight (and which can thus be directly compared to the experimental ones), are in general not available at low temperatures.
We refer to the literature for further details~\cite{gubernatis:2016}.
We made use of the ALF libraries, which use the jackknife resampling scheme for error estimation~\cite{assaad:2022,assaad:2022a}.

\subsection{Thermometry}

The thermometry procedure relies on comparing experimental and dQMC spin correlations.
The dQMC simulations cover a range of doping $|\delta| \leq 0.3$ and temperatures $T \gtrsim t/4$, except at half-filling where the absence of the sign problem allows to go to much lower temperatures.
In practice, the dQMC calculations are performed on a square system of size $L\times L$ with periodic boundary conditions, with $L = \num{16}$ at half-filling and $L = \num{10}$ otherwise.

The finite precision of dQMC calculations are taken into account by determining the maximum distance $\dmax$ below which dQMC spin correlations are significantly non-zero, \emph{i.e.} $|C_{\rm ss}^{(2)}(|\bm d| \leq \dmax)| > 0$ (including errorbars).
The comparison is then performed only considering the data points for which $|\bm d| \leq \dmax$.
In practice, we also impose $\sqrt{2} \leq \dmax \leq L/2$.

We interpolate linearly the dQMC data between settings $(\delta, T)$ where the calculation is actually performed, and we find the best fit between dQMC simulations and experimental data.
The temperature $T$ is the only free parameter of the fitting procedure, with the doping level $\delta$, directly measured in the experimental snapshots, being fixed.
We report in \figref{fig:figS5} the 119 datasets that are used in this work, with the corresponding thermometry fit.

\subsubsection*{Extrapolation to lower temperatures}

\begin{figure}[!t]
\centering
\includegraphics[scale=1]{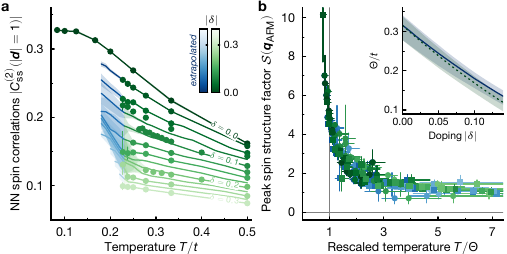}
\caption{
\textbf{Effect of extrapolation at low temperature.}
\textbf{a.} NN spin correlations in the dQMC simulations (disks) as a function of temperature and doping, and results of the 2D linear interpolation (lines) for a chosen set of doping values.
Note that the dQMC calculations are run at fixed chemical potential in the grand canonical ensemble, and the resulting doping may not correspond to the rounded values of the lines.
The presence of low-temperature dQMC data at half-filling allows to extrapolate finite doping data to lower temperatures (blue lines).
\textbf{b.} Result of the collapse procedure with (blue) and without (green) the extrapolated data.
Note that the blue data here is identical to \figref{fig:fig3}d.
The inset shows the resulting temperature scale $\scaling$, with (blue line) and without (green dashed line) using the extrapolated data from panel (\textbf{a}).
}
\label{fig:figS7}
\end{figure}

Because dQMC data extends to lower temperatures at half-filling, the interpolation procedure allows to \emph{extrapolate} dQMC spin correlations to temperatures $T \lesssim t/4$ at finite doping.
The reported temperatures below $t/4$, in datasets with finite doping, are thus obtained from this extrapolation.
In \figref{fig:figS7}, we show that the extrapolation has a negligible effect on the reported scaling temperature scale $\scaling$: when omitting the datasets with extrapolated temperatures from the collapse optimization, the resulting $\scaling(\delta)$ remains very close to what is presented in \figref{fig:fig3}.

\begin{figure}[!t]
\centering
\includegraphics[scale=1]{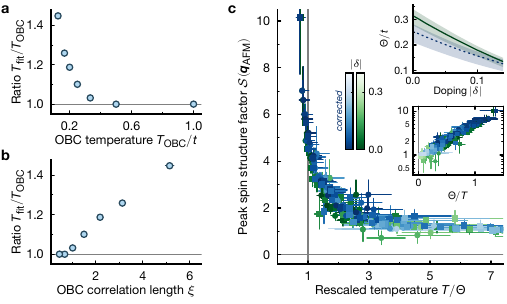}
\caption{
\textbf{Effect of boundary condition corrections.}
\textbf{a.} Result of the fitting procedure on OBC dQMC simulations at half-filling.
For $T_{\rm OBC} \lesssim 0.3$, significant corrections are expected to emerge.
\textbf{b.} Same data as in panel (\textbf{a}), plotted vs. the OBC correlation length $\xi$.
The correction to larger doping is deduced from these results.
\textbf{c.} Result of the collapse procedure with (blue) and without (green) the boundary condition correction.
Note that the green data here is identical to \figref{fig:fig3}d.
The lower inset shows the same data in semi-log scale.
The upper inset shows the resulting temperature scale $\scaling$, with (blue dashed line) and without (green line) the boundary condition correction (\textbf{a}).
}
\label{fig:figS8}
\end{figure}
\subsubsection*{Finite-size effects in dQMC}

Spin correlations will naturally be affected by the system size or the boundary conditions of the simulation whenever the correlation length becomes sizeable.
We quantify this effect by comparing open and periodic boundary conditions (OBC and PBC, respectively) at half-filling in the dQMC simulation.
More precisely, we fit the OBC calculations using the same fitting procedure as for the experimental data, \emph{i.e.} using the PBC simulations, and show the results in \figref{fig:figS8}a-b.
We find that, for temperatures $T/t \lesssim 0.3$, deviations arise between the true temperature $T_{\rm OBC}$ and the fitted temperature $T_{\rm fit}$, corresponding to correlation lengths of about 1-2 sites.
Interestingly, we always find $T_{\rm fit} > T_{\rm OBC}$; this is expected from the fact that PBC tend to increase spin correlations at larger distances, meaning that a larger temperature is sufficient to match the OBC spin correlations.
As a consequence, the experimental temperatures reported in this work remain conservative, in the sense that they are potentially biased towards larger values.

We evaluate the effect of this bias in our estimation of the temperature scale $\scaling$, by applying a correlation-length-dependent correction to the temperature deduced from the results of \figref{fig:figS8}b.
We show in \figref{fig:figS8}c the result of the collapse procedure, with (blue) and without (green) the correction.
The results remain quantitatively the same, although the extracted $\scaling(\delta)$ shows a slightly smaller value close to half-filling.

\subsubsection*{Effects of spatial inhomogeneities}

\begin{figure*}[!t]
\centering
\includegraphics[scale=1]{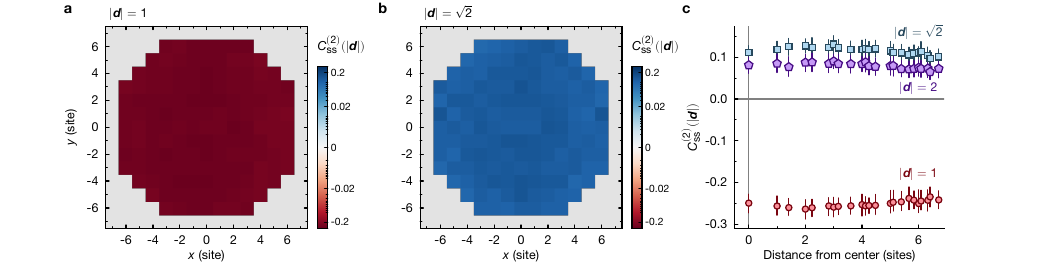}
\caption{
\textbf{Local spin correlations.}
\textbf{a, b.} Local NN (\textbf{a}) and NNN (\textbf{b}) spin correlations across the system $\Omega$.
The colourscale is the same as in \figref{fig:fig3}a.
\textbf{c.} Radially averaged correlations as a function of the distance from the trap center.}
\label{fig:figS6}
\end{figure*}

\begin{figure*}[!thpb]
\centering
\includegraphics[scale=1]{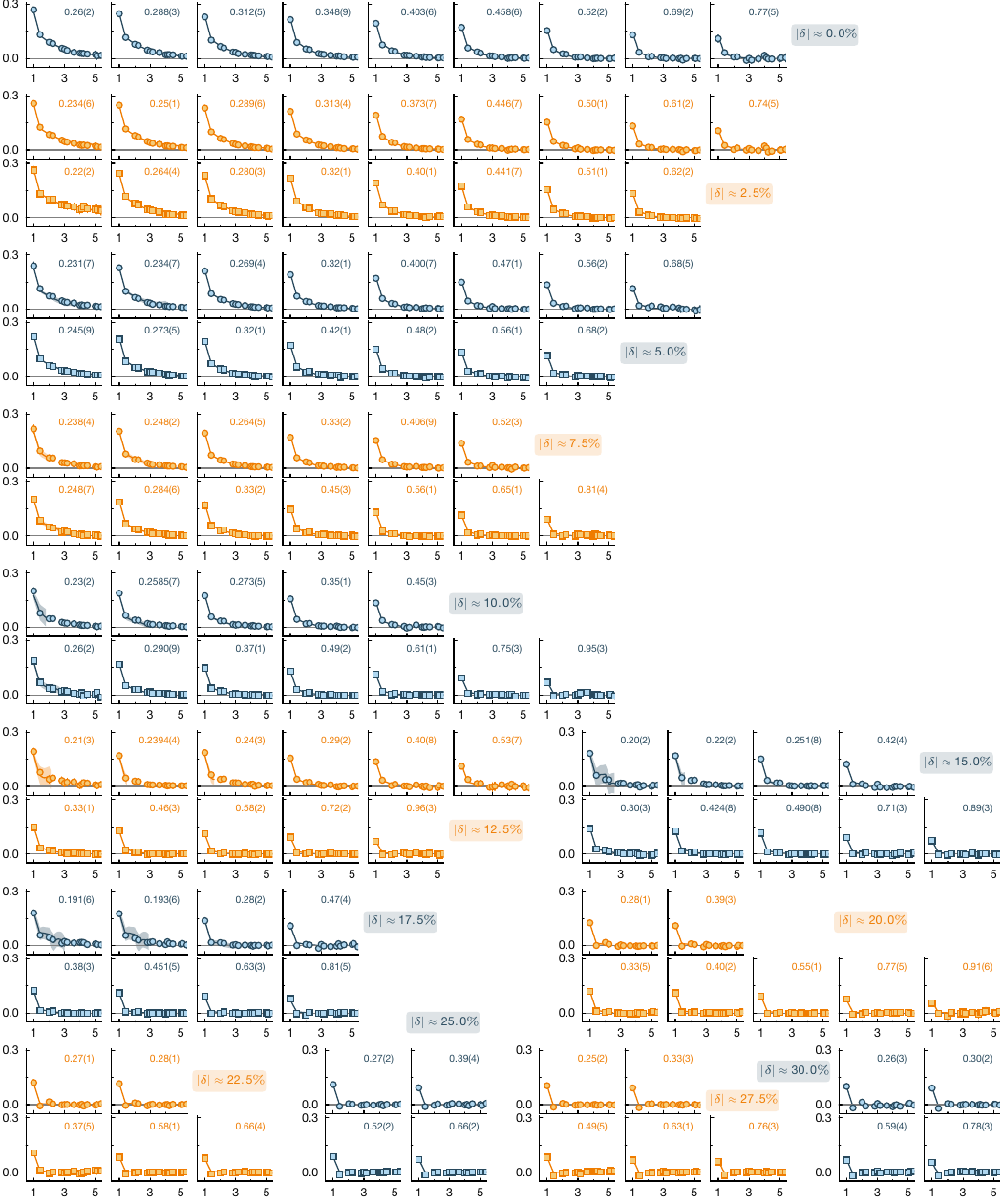}
\caption{
\textbf{Spin correlations for thermometry.}
Rectified spin correlations vs. distance for each dataset used in this work.
The disks and squares are the experimental hole- and doublon-doped data, respectively.
The solid line is the resulting dQMC fit from which the temperature is extracted, also indicated on each panel.
For readability, we omitted the axis labels (see \figref{fig:fig2} for reference).
}
\label{fig:figS5}
\end{figure*}

We check in \figref{fig:figS6} the effects of the trapping potential on the spin correlations.
In particular, we show the local NN and NNN spin correlations across the system $\Omega$ at half-filling, as well as a radial average from the trap center.
We do not measure any significant spatial inhomogeneities, indicating that the trapping potential has a negligible effect on the thermometry.

\subsection{Minimally entangled thermal typical states (METTS)}

Thermal expectation values at inverse temperature $\beta$ of an operator $\hat{\mathcal{O}}$ can, in general, be evaluated as,
\begin{align}
\braket{\hat{\mathcal{O}}}_\beta &= \frac{1}{\mathcal{Z}} \text{Tr}(\ee^{-\beta H} \hat{\mathcal{O}}) \nonumber\\
&= \frac{1}{\mathcal{Z}} \sum_i  \braket{\sigma_i | \ee^{-\beta \hat H/2} \hat{\mathcal{O}} \ee^{-\beta \hat H/2} | \sigma_i}\nonumber\\
&=\frac{1}{\mathcal{Z}} \sum_i p_i \braket{\psi_i | \hat{\mathcal{O}} | \psi_i}.
\label{eq:metts_thermal_average}
\end{align}
Here, $\ket{\sigma_i}$ denotes the basis of classical product states,  
\begin{equation}
    p_i= \braket{\sigma_i | \ee^{-\beta \hat H} | \sigma_i}\geq 0,
\end{equation}
and $\mathcal{Z} = \sum_i p_i$ is the partition function. The pure quantum states,
\begin{equation}
\ket{\psi_i} = \frac{1}{\sqrt{p_i}}\ee^{-\beta \hat H/2} \ket{\sigma_i}, 
\label{eq:metts_thermal_evol}
\end{equation}
are referred to as the \textit{minimally entangled typical thermal states} (METTS).
The METTS algorithm samples the METTS states $\ket{\psi_i}$ by constructing a Markov chain whose stationary distribution is given by the probabilities $p_i/\mathcal{Z}$, cf. refs.~\cite{white:2009,stoudenmire:2010}.
The weights $p_i \geq 0$ are nonnegative real numbers and, hence, no sign problem is encountered.
The challenge lies in computing the imaginary-time evolutions of the product states $\ket{\sigma_i}$.
Recent advances in time-evolution methods for matrix product states allow accurate computations of the METTS states $\ket{\psi_i}$~\cite{paeckel:2019}.
In our simulations, we mostly employ the time-dependent variational principle~\cite{haegeman:2011,haegeman:2016}, with subspace expansion~\cite{yang:2020b}, to perform this time evolution.
Details of our simulations are described in detail in Ref.~\cite{wietek:2021}. We used a $32 \times 4$ cylinder for the magnetic susceptibility calculations and a $16 \times 4$ cylinder for the higher-order correlators.
Bond dimensions of up to $D=2500$ were used to reach convergence within the desired doping and temperature range.
Error estimates are obtained by standard means of error analysis of Markov chain Monte Carlo, i.e. from estimation of the standard error of the mean, an estimate of autocorrelation times, and monitoring of thermalisation.
We refer to Ref.~\cite{wietek:2021} for further details.

\subsection{Classical snapshots from METTS}
While conventionally, expectation values in the METTS algorithm are evaluated by computing expectation values of the form $\braket{\psi_i | \hat{\mathcal{O}} | \psi_i}$, METTS can also be employed in a mode that is very similar to the measurements performed in the quantum gas microscope.
Within METTS, the classical product states $\ket{\sigma_i}$ are sampled with probability $p_i/\mathcal{Z} = \braket{\sigma_i | \ee^{-\beta \hat H} | \sigma_i} / \mathcal{Z}$.
This probability is exactly the probability of sampling a product state from the thermal density matrix $\rho = \ee^{-\beta \hat H}$ according to the Born rule of probability.
As such, the classical snapshots $\ket{\sigma_i}$ from METTS can be compared one-to-one to the snapshots obtained from the quantum gas microscope.
Importantly, observables $\hat{\mathcal{D}}$ which are diagonal in the computational basis $\ket{\sigma_i}$, \emph{i.e.} $\hat{\mathcal{D}}\ket{\sigma_i} = d_i\ket{\sigma_i}$, where $d_i \in \mathbb{C}$ for all $\ket{\sigma_i}$, can be evaluated directly on the classical snapshots.
This follows from, 
\begin{align}
\begin{split}
\braket{\hat{\mathcal{D}}} &= \frac{1}{\mathcal{Z}} \sum_i \braket{\sigma_i | \ee^{-\beta \hat H} \hat{\mathcal{D}} | \sigma_i} =
\frac{1}{\mathcal{Z}} \sum_i d_i \braket{\sigma_i | \ee^{-\beta \hat H} | \sigma_i}  \\
&= \frac{1}{\mathcal{Z}} \sum_i p_i  d_i = \frac{1}{\mathcal{Z}} \sum_i p_i \braket{\sigma_i |\hat{\mathcal{D}}| \sigma_i}.
\end{split}
\end{align}
This allows to evaluate observables from snapshots obtained by the experimental as well as by METTS using the same data analysis scripts.

\subsection{Geometric String Calculations}
The geometric string picture constitutes an effective microscopic description of mobile dopants in a surrounding antiferromagnet.
It is based on a detailed modeling how a single doped hole distorts its spin background and was originally conceived to capture magnetic polaron formation~\cite{grusdt:2018a}.
Here we consider its simplest extension to finite doping, where correlations between different dopants are neglected, however strong spin-charge correlations arise from how every individual dopant distorts the originally undoped spin background. 

The main assumption of the geometric string picture is the so-called frozen spin approximation.
It assumes that the motion of the dopant only displaces the surrounding spins along its trajectory, while keeping their spin state and its entanglement with the surrounding spins unchanged.
Furthermore it is assumed that any distinct trajectories $\Sigma$ and $\Sigma'$, defined without self-retracing paths and defining strings on the square lattice, correspond to orthonormal states $\braket{\Sigma | \Sigma'} = \delta_{\Sigma,\Sigma'}$.
This leads to an effective hopping Hamiltonian on a Bethe lattice, where each site corresponds to a given string configuration $\ket{\Sigma}$ and encodes strong spin-charge correlations through the underlying frozen-spin approximation.
For a single dopant moving in a square lattice AFM, the geometric string picture has been shown to be very accurate~\cite{grusdt:2019}.
This picture goes back to early work on the magnetic polaron problem in an Ising AFM~\cite{bulaevskii:1968} and has continuously been developed further~\cite{shraiman:1988,trugman:1988,beran:1996,grusdt:2018a,bermes:2024}. 

\begin{figure}[!t]
\centering
\includegraphics[scale=1]{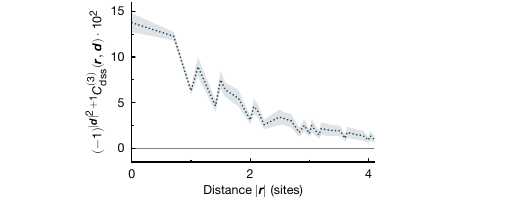}
\caption{
\textbf{Dopant-spin-spin correlations in geometric spin theory.}
Sign corrected three-point correlators as a function of distance, for a single dopant evolving in a frozen AFM background characterized by a fixed two-point spin-spin correlation $C_{\mathrm{ss}}^{(2)}(\bm d)$.
See text for additional details.
}
\label{fig:figS4}
\end{figure}

Here we apply the geometric string theory in two forms.
First, in \figref{fig:figS4}, we computed the three-point spin-spin-charge correlations expected around a single dopant.
To this end we assumed a frozen spin background characterized by spin-spin correlations $C_{\mathrm{ss}}^{(2)}(\bm{d})$ and ignoring the effects of virtual doublon-hole pairs.
Then we computed the thermal state of fluctuating strings on the Bethe lattice, at the lowest experimental temperature, and sampled string configurations from it.
Each string configuration corresponds to a particular re-shuffling of the surrounding spins, which allows us to compute dopant-spin-spin correlations in \figref{fig:figS4}.
The emergent polaron tail reflects the sizeable probability, facilitated by both thermal and quantum fluctuations, of finding strings extending over several lattice sites.

Second, in Figs. \ref{fig:fig4}b and \ref{fig:fig5} of the main text, we compute $3^{\mathrm{th}}$, $4^{\mathrm{th}}$ and $5^{\mathrm{th}}$ order correlators at various doping values.
To this end we followed the procedure introduced in~\cite{chiu:2019a}, i.e. we start from experimental snapshots at zero doping.
To produce string theory snapshots at finite doping, we first randomly remove spins to introduce non-zero doping.
For each dopant, we sample one string configuration from the thermal string state on the Bethe lattice and move the dopant along this string, displacing spins along the way.
At each doping value, the analysis of the final string data set proceeds identical to the analysis of the experimental snapshots. 

The comparison of \figref{fig:fig4}b in the main text and \figref{fig:figS4} shows that the effects of finite doping and doublon-hole pairs lead to a slightly shorter polaronic tail.
In general, the geometric string approach assumes strong coupling and is expected to become less reliable when more doublon-hole pairs appear.

\subsection{Spin structure factor and magnetic susceptibility}

We define the spin structure factor $\mathcal{S}(\bm q)$ in the text with Eq. \eqref{eq:Sq}, and compute it experimentally by restricting the spin correlation functions $C_{\mathrm{ss}}^{(2)}(\bm d)$ to distances $|\bm d| \leq 7$ sites.
We assume that the peak structure factor $\Spp$ follows a temperature scaling of the form given by Eq. \eqref{eq:weak_coupling}, $\Spp = A(\ee^{2\scaling(\delta)/T} - 1) + 1$.
In the present work, we establish that the temperature scale $\scaling$ decreases with doping, and matches the temperature $T^*$ for the pseudogap (\figref{fig:fig3}).
As explained in the main text, we determine experimentally $\scaling(\delta)$ through a second order expansion in $|\delta|$, \emph{i.e.} by writing $\scaling(\delta) = \scaling^{(0)} + \scaling^{(1)}|\delta| + \scaling^{(2)}|\delta|^{2}$ and treating $\scaling^{(i)}$ ($i = 0, 1, 2$) and $A$ as fit parameters.

We first fit the the coefficient $\scaling^{(0)}$ to the $|\delta| = 0$ data (half-filling).
In the following steps, this coefficient is fixed.
Then, for a given $(\scaling^{(1)}, \scaling^{(2)})$, we fit the scaling function of Eq. \eqref{eq:weak_coupling} to the curve $\Spp(\scaling/T)$ that we compute by combining all the datasets.
This fit gives a reduced $\chi^{2}$, which we use as a ``goodness-of-fit''.
The procedure is iterated to find the set of parameters $(\scaling^{(1)}, \scaling^{(2)})$ that minimize $\chi^{2}$, \emph{i.e.} that optimize the collapse of $\Spp(\delta, T)$ onto a universal curve.

The magnetic susceptibility $\chi_m$ shows a maximum at the pseudogap temperature $T = T^*$.
This is characteristic of the emergence of the pseudogap, and constituted the first experimental observation of the pseudogap in cuprates with large local moments \cite{johnston:1989}.
Formally, the magnetic susceptibility can be deduced from the spin structure factor as $T\chi_m = |\Omega|\mathcal{S}(\bm{q} = \bm{0})$ \cite{hartke:2023}.
Experimentally, we extract the low-$\bm{q}$ behaviour of the spin structure factor $\mathcal{S}(\bm{q})$ using a quadratic fit in the region $|\bm q| \leq \pi/4$ with a symmetric quadratic form and offset, \emph{i.e.} $\mathcal{S}(\bm q) \approx c_0 + c_1(q_x^{2} + q_y^{2}) + c_2q_xq_y$, and use resulting fitted offset as $\mathcal{S}(\bm q = \bm 0)$.
We find that this procedure yields a less noisy result compared to simply summing the real space correlations $C_{\mathrm{ss}}^{(2)}(\bm d)$ over the analysis region $\Phi$, especially at larger temperature or doping where the extension of spin correlations is very limited.

\begin{figure}[!t]
\centering
\includegraphics[scale=1]{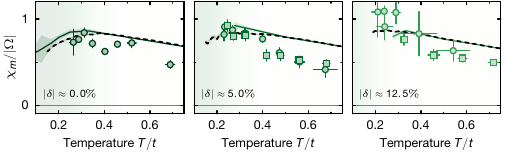}
\caption{
\textbf{Magnetic susceptibility.}
Magnetic susceptibility $\chi_m/|\Omega|$ as a function of temperature for half-filling ($|\delta| \approx \SI{0}{\percent}$) and at larger doping ($|\delta| \approx \SI{5}{\percent}$ and $\SI{12.5}{\percent}$).
Disks (squares) represent hole-doped and half-filled (doublon-doped) experimental data.
Solid lines are obtained from dQMC simulations (down to $T/t \approx 0.25$), and dashed lines correspond to METTS simulations.
The shaded region corresponds to $T \lesssim \scaling$.
}
\label{fig:figS9}
\end{figure}

We show the resulting spin susceptibility in \figref{fig:figS9}, as well as numerical values from dQMC (solid lines) and METTS (dashed lines).
We find notable discrepancies between experimental data and numerical results, especially at larger temperature.
Nevertheless, our data is still compatible with the appearance of a plateau of the susceptibility at low temperature, and corroborates the emergence of a pseudogap.
In practice, the METTS simulations suggest that maximum of susceptibility is quite broad, meaning that an unambiguous measurement of the downturn would require going to much lower temperatures.

\subsection{Dopant-spin-spin correlations}

\begin{figure*}[!thpb]
\centering
\includegraphics[scale=1]{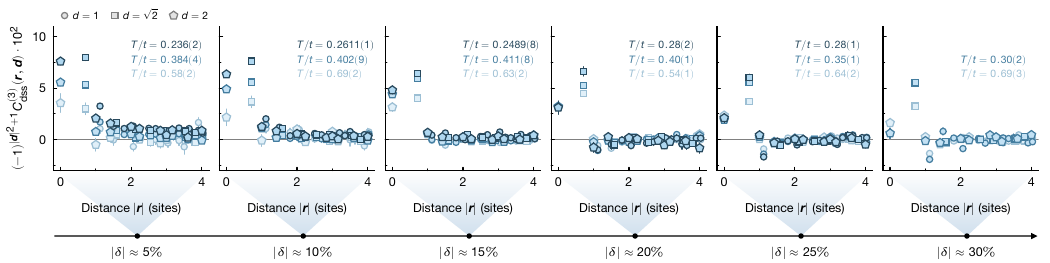}
\caption{
\textbf{Temperature and doping evolution of dopant-spin-spin correlations.}
Experimental sign-corrected dopant-spin-spin correlations as a function of the distance $|\bm r|$ between dopant and spin bonds.
The colour indicates the temperature, and is labelled in each panel.
The doping level goes from \SI{5}{\percent} to \SI{30}{\percent} from left to right.
}
\label{fig:figS2}
\end{figure*}

For the sake of clarity, we only show in \figref{fig:fig4} the low temperature shape and strength of the three-point dopant-spin-spin correlations that characterize magnetic polarons.
We compare in \figref{fig:figS2} the experimental data corresponding to different temperatures and doping levels and recover the general features mentioned in the main text.
Specifically, the extended tail only forms in the low temperature and low doping regime characterized by $T < \Delta(\delta) \simeq T^*$.
The temperature dependence of the polaron core ($|\bm r| \lesssim 1.5$) is important at lower doping, and washes out at larger doping where the physics is dominated by Fermi liquid behaviour.

\subsection{Connected and conditional correlations}

Connected correlations are defined as a multi-variate cumulant \cite{kubo:1962},
\begin{equation}
    \braket{\hat X_1\cdots\hat X_n}_\mathrm{c} = \left.\frac{\partial}{\partial z_1} \cdots \frac{\partial}{\partial z_n}\log\braket{\ee^{\sum_i z_i \hat X_i}}\right|_{z_i = 0}.
\end{equation}
Orders 1 and 2 correspond to the mean ($\braket{\hat X}_\mathrm{c} = \braket{\hat X}$) and covariance ($\braket{\hat X_1 \hat X_2}_\mathrm{c} = \braket{\hat X_1 \hat X_2} - \braket{\hat X_1}\braket{\hat X_2}$), respectively.
More generally, the connected correlation of order $n$ consists of the bare correlation $\braket{\prod_{i=1}^{n}\hat X_i}$ to which is subtracted the connected correlations of all possible lower order terms.

Note that all the correlation functions presented in this work are averaged throughout the system --- this is the meaning of the sums in Eqs. \eqref{eq:Css}, \eqref{eq:polaron}, \eqref{eq:4point} and \eqref{eq:5point}.
For each of these equations, we introduce a normalization constant that counts the number of configurations within the system.

We give below explicit expression for order $n = 3, 4$.
The $5^{\mathrm{th}}$ order expression (not shown here) is built in a similar fashion, and contains more terms.
\begin{widetext}
\begin{align}
\braket{\hat{A}\hat{B}\hat{C}}_\mathrm{c} & = \braket{\hat{A}\hat{B}\hat{C}} - \braket{\hat{A}}\braket{\hat{B}\hat{C}}_\mathrm{c} - \braket{\hat{B}}\braket{\hat{C}\hat{A}}_\mathrm{c} - \braket{\hat{C}}\braket{\hat{A}\hat{B}}_\mathrm{c} - \braket{\hat{A}}\braket{\hat{B}}\braket{\hat{C}} \nonumber \\[.7em]
& = \braket{\hat{A}\hat{B}\hat{C}} - \braket{\hat{A}}\braket{\hat{B}\hat{C}} - \braket{\hat{B}}\braket{\hat{C}\hat{A}} - \braket{\hat{C}}\braket{\hat{A}\hat{B}} + 2\braket{\hat{A}}\braket{\hat{B}}\braket{\hat{C}} \\[1em]
\braket{\hat{A}\hat{B}\hat{C}\hat{D}}_\mathrm{c} & = \braket{\hat{A}\hat{B}\hat{C}\hat{D}} - \braket{\hat{A}}\braket{\hat{B}\hat{C}\hat{D}}_\mathrm{c} - \braket{\hat{B}}\braket{\hat{C}\hat{D}\hat{A}}_\mathrm{c} - \braket{\hat{C}}\braket{\hat{D}\hat{A}\hat{B}}_\mathrm{c} - \braket{\hat{D}}\braket{\hat{A}\hat{B}\hat{C}}_\mathrm{c} \nonumber \\
    & - \braket{\hat{A}\hat{B}}_\mathrm{c}\braket{\hat{C}\hat{D}}_\mathrm{c} - \braket{\hat{A}\hat{C}}_\mathrm{c}\braket{\hat{B}\hat{D}}_\mathrm{c} - \braket{\hat{A}\hat{D}}_\mathrm{c}\braket{\hat{B}\hat{C}}_\mathrm{c} \nonumber \\
    & - \braket{\hat{A}\hat{B}}_\mathrm{c}\braket{\hat{C}}\braket{\hat{D}} - \braket{\hat{A}\hat{C}}_\mathrm{c}\braket{\hat{B}}\braket{\hat{D}} - \braket{\hat{A}\hat{D}}_\mathrm{c}\braket{\hat{B}}\braket{\hat{C}} - \braket{\hat{B}\hat{C}}_\mathrm{c}\braket{\hat{A}}\braket{\hat{D}} - \braket{\hat{B}\hat{D}}_\mathrm{c}\braket{\hat{A}}\braket{\hat{C}} - \braket{\hat{C}\hat{D}}_\mathrm{c}\braket{\hat{A}}\braket{\hat{B}} \nonumber \\
    & - \braket{\hat{A}}\braket{\hat{B}}\braket{\hat{C}}\braket{\hat{D}} \nonumber \\[.7em]
& = \braket{\hat{A}\hat{B}\hat{C}\hat{D}} - \braket{\hat{A}}\braket{\hat{B}\hat{C}\hat{D}} - \braket{\hat{B}}\braket{\hat{C}\hat{D}\hat{A}} - \braket{\hat{C}}\braket{\hat{D}\hat{A}\hat{B}} - \braket{\hat{D}}\braket{\hat{A}\hat{B}\hat{C}} \nonumber \\
    & - \braket{\hat{A}\hat{B}}\braket{\hat{C}\hat{D}} - \braket{\hat{A}\hat{C}}\braket{\hat{B}\hat{D}} - \braket{\hat{A}\hat{D}}\braket{\hat{B}\hat{C}} \nonumber \\
    & +2 \left[\braket{\hat{A}\hat{B}}\braket{\hat{C}}\braket{\hat{D}} + \braket{\hat{A}\hat{C}}\braket{\hat{B}}\braket{\hat{D}} + \braket{\hat{A}\hat{D}}\braket{\hat{B}}\braket{\hat{C}} + \braket{\hat{B}\hat{C}}\braket{\hat{A}}\braket{\hat{D}} + \braket{\hat{B}\hat{D}}\braket{\hat{A}}\braket{\hat{C}} + \braket{\hat{C}\hat{D}}\braket{\hat{A}}\braket{\hat{B}}\right] \nonumber \\
    & - 6\braket{\hat{A}}\braket{\hat{B}}\braket{\hat{C}}\braket{\hat{D}}
\end{align}
\end{widetext}

Following the logic in Ref. \cite{bohrdt:2021b}, we chose the conditional correlation function used in \figref{fig:fig5} as to capture the connected $4^{\mathrm{th}}$-order spin correlation on the condition of the presence of a dopant in the vicinity.
This quantity is effectively a $5^{\mathrm{th}}$-order correlation, although not a \emph{connected} correlation per se.
Its exact expression is
\begin{widetext}
\begin{multline}
    \left.C_{\mathcal{K}}^{(5)}(\hat d_0, \hat S_1^z, \hat S_2^z, \hat S_3^z, \hat S_4^z)\right|_{\mathrm{cond.}\ \hat d_0} = 2^4 \times \left[
        \frac{\braket{\hat d_0\hat S_1^z\hat S_2^z\hat S_3^z\hat S_4^z}}{\braket{\hat d_0}}\right. \\
            - \frac{\braket{\hat d_0\hat S_1^z\hat S_2^z\hat S_3^z}}{\braket{\hat d_0}}\frac{\braket{\hat d_0\hat S_4^z}}{\braket{\hat d_0}}
            - \frac{\braket{\hat d_0\hat S_1^z\hat S_2^z\hat S_4^z}}{\braket{\hat d_0}}\frac{\braket{\hat d_0\hat S_3^z}}{\braket{\hat d_0}}
            - \frac{\braket{\hat d_0\hat S_1^z\hat S_3^z\hat S_4^z}}{\braket{\hat d_0}}\frac{\braket{\hat d_0\hat S_2^z}}{\braket{\hat d_0}}
            - \frac{\braket{\hat d_0\hat S_2^z\hat S_3^z\hat S_4^z}}{\braket{\hat d_0}}\frac{\braket{\hat d_0\hat S_1^z}}{\braket{\hat d_0}} \\
- \frac{\braket{\hat d_0\hat S_1^z\hat S_2^z}}{\braket{\hat d_0}}\frac{\braket{\hat d_0\hat S_3^z\hat S_4^z}}{\braket{\hat d_0}}
            - \frac{\braket{\hat d_0\hat S_1^z\hat S_3^z}}{\braket{\hat d_0}}\frac{\braket{\hat d_0\hat S_2^z\hat S_4^z}}{\braket{\hat d_0}}
            - \frac{\braket{\hat d_0\hat S_1^z\hat S_4^z}}{\braket{\hat d_0}}\frac{\braket{\hat d_0\hat S_2^z\hat S_3^z}}{\braket{\hat d_0}} \\
- \frac{\braket{\hat d_0\hat S_1^z\hat S_2^z}}{\braket{\hat d_0}}\frac{\braket{\hat d_0\hat S_3^z}}{\braket{\hat d_0}}\frac{\braket{\hat d_0\hat S_4^z}}{\braket{\hat d_0}}
            - \frac{\braket{\hat d_0\hat S_1^z\hat S_3^z}}{\braket{\hat d_0}}\frac{\braket{\hat d_0\hat S_2^z}}{\braket{\hat d_0}}\frac{\braket{\hat d_0\hat S_4^z}}{\braket{\hat d_0}}
            - \frac{\braket{\hat d_0\hat S_1^z\hat S_4^z}}{\braket{\hat d_0}}\frac{\braket{\hat d_0\hat S_2^z}}{\braket{\hat d_0}}\frac{\braket{\hat d_0\hat S_3^z}}{\braket{\hat d_0}} \\
            - \frac{\braket{\hat d_0\hat S_2^z\hat S_3^z}}{\braket{\hat d_0}}\frac{\braket{\hat d_0\hat S_1^z}}{\braket{\hat d_0}}\frac{\braket{\hat d_0\hat S_4^z}}{\braket{\hat d_0}}
            - \frac{\braket{\hat d_0\hat S_2^z\hat S_4^z}}{\braket{\hat d_0}}\frac{\braket{\hat d_0\hat S_1^z}}{\braket{\hat d_0}}\frac{\braket{\hat d_0\hat S_3^z}}{\braket{\hat d_0}}
            - \frac{\braket{\hat d_0\hat S_3^z\hat S_4^z}}{\braket{\hat d_0}}\frac{\braket{\hat d_0\hat S_1^z}}{\braket{\hat d_0}}\frac{\braket{\hat d_0\hat S_2^z}}{\braket{\hat d_0}} \\
\left.- \frac{\braket{\hat d_0\hat S_1^z}}{\braket{\hat d_0}}\frac{\braket{\hat d_0\hat S_2^z}}{\braket{\hat d_0}}\frac{\braket{\hat d_0\hat S_3^z}}{\braket{\hat d_0}}\frac{\braket{\hat d_0\hat S_4^z}}{\braket{\hat d_0}}
        \right],
\end{multline}
\end{widetext}
although all the terms of the form $\braket{\hat d_0 \hat S_i^z}$ or $\braket{\hat d_0 \hat S_i^z\hat S_j^z\hat S_k^z}$ containing an odd number of spin operators vanish by symmetry.

\subsubsection*{Additional data for $4^{\rm th}$- and $5^{\rm th}$-order correlators}

\begin{figure}[!thpb]
\centering
\includegraphics[scale=1]{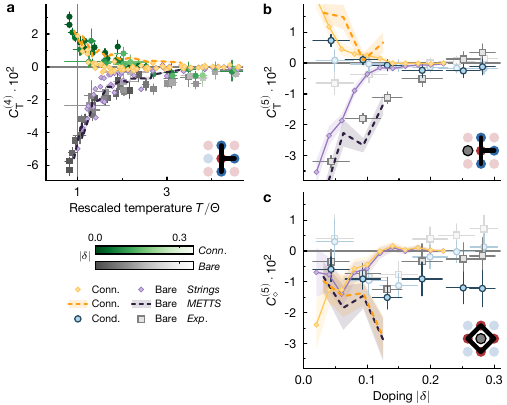}
\caption{
\textbf{Higher-order correlations --- extended data.}
\textbf{a} $4^\mathrm{th}$-order spin correlations with 4 spins arranged in a in a T-shape.
The connected (bare) correlations, represented by green circles (grey squares), are plotted against the rescaled temperature $T/\scaling$.
The colour scale indicates the doping level.
\textbf{b, c} $5^\mathrm{th}$-order spin-charge correlations with 4 spins arranged in a T-shape close to a dopant (\textbf{b}) or in a diamond shape (\textbf{c}) around the dopant.
The conditioned correlations (blue circles) are compared to the bare correlations (grey squares), and plotted against doping.
Lighter colours indicate higher temperatures.
In all panels, the dashed lines and the dots correspond to METTS and geometric string calculations, respectively.
}
\label{fig:figS10}
\end{figure}

We give in \figref{fig:figS10}a,b the $4^{\rm th}$- and $5^{\rm th}$-order correlations in the case of a T-shape configuration of spins.
Figure~\ref{fig:figS10}c shows the $5^{\rm th}$-order correlation in a diamond configuration for two different temperatures $T/t \approx 0.25$ and $T/t \approx 0.6$ (the lowest temperature, corresponding to the darker points, is identical to the data in \figref{fig:fig5}c).

We recover in $C_T^{(4)}$ the main features found in $C_\Box^{(4)}$, namely non-zero connected correlations in the regime $T \lesssim \scaling$.
Compared to the square and the diamond configurations (\figref{fig:fig5}a,b), the signs are inverted, as expected from the AFM pattern.

For the $5^{\rm th}$-order T-shape configuration, the conditioned correlations are non-vanishing only for the coldest dataset, close to half-filling (dark blue points in \figref{fig:figS10}b)
The bare correlations (grey squares), on the other hand, are negative and increase in amplitude as doping or temperature is reduced, following a similar trend as the bare $C_T^{(4)}$.
Interestingly, $C_\diamond^{(5)}$ shows little dependence to temperature, as one recovers similar features of dominant higher correlations around $\num{10}-\SI{20}{\percent}$ doping.

\begin{thebibliography}{100}
\bibitem{note:symmetry}
\bibinfo{note}{The particle-hole symmetry here means that hole- and doublon-doped exhibit the exact same physical properties. In practice, we found experimentally that we can reach lower temperatures in hole-doped systems. We exploit this symmetry in evaluating higher-order correlators (sections~\ref{sec:secIII} and \ref{sec:secIV}) by mixing both hole-doped and doublon-doped sectors, thus improving statistics (see SI).}
\end{thebibliography}
\end{document}